\documentclass[conference]{IEEEtran}
\IEEEoverridecommandlockouts
\usepackage{cite}
\usepackage{amsmath,amssymb,amsfonts}
\usepackage{graphicx}
\usepackage{textcomp}
\usepackage[table]{xcolor}
\usepackage{subcaption}
\usepackage{balance}
\usepackage[hyphens]{url}
\usepackage[hidelinks]{hyperref}
\usepackage{tabularx}
\usepackage{listings}
\usepackage{enumitem}
\usepackage{xcolor}
\usepackage{multirow}
\usepackage{float}
\usepackage{stfloats}
\usepackage{orcidlink}

\definecolor{codegreen}{rgb}{0,0.6,0}
\definecolor{codegray}{rgb}{0.5,0.5,0.5}
\definecolor{codepurple}{rgb}{0.58,0,0.82}
\definecolor{backcolour}{rgb}{0.95,0.95,0.92}
\definecolor{nvidiacolor}{rgb}{0.4627, 0.7255, 0}
\definecolor{lightred}{rgb}{0.9255, 0.5176, 0.5176}

\lstdefinestyle{mystyle}{
    backgroundcolor=\color{backcolour},   
    commentstyle=\color{codegreen},
    keywordstyle=\color{magenta},
    numberstyle=\tiny\color{codegray},
    stringstyle=\color{codepurple},
    basicstyle=\ttfamily\footnotesize,
    breakatwhitespace=false,         
    breaklines=true,                 
    captionpos=b,                    
    keepspaces=true,                 
    numbers=left,                    
    numbersep=5pt,                  
    showspaces=false,                
    showstringspaces=false,
    showtabs=false,                  
    tabsize=2,
    xleftmargin=1.2em
}

\def\BibTeX{{\rm B\kern-.05em{\sc i\kern-.025em b}\kern-.08em
    T\kern-.1667em\lower.7ex\hbox{E}\kern-.125emX}}

\newcommand{\nvidia}{NVIDIA}
\newcommand{\holoscan}{Holoscan}
\newcommand{\nvidiaholoscan}{NVIDIA Holoscan}
\newcommand{\holoscansdk}{NVIDIA Holoscan SDK}

\begin{document}

\title{Towards Deterministic End-to-end Latency for Medical AI Systems in 
\nvidiaholoscan{}}

\author{\IEEEauthorblockN{Soham Sinha \orcidlink{0000-0001-8962-4714}}
	\IEEEauthorblockA{\textit{NVIDIA} \\
		Santa Clara, CA, USA\\
		sohams@nvidia.com}
	\and
	\IEEEauthorblockN{Shekhar Dwivedi}
	\IEEEauthorblockA{\textit{NVIDIA} \\
		Santa Clara, CA, USA\\
		shekhard@nvidia.com}
	\and
	\IEEEauthorblockN{Mahdi Azizian}
	\IEEEauthorblockA{\textit{NVIDIA} \\
		Santa Clara, CA, USA\\
		mazizian@nvidia.com}
}

\maketitle

\begin{abstract}
The introduction of AI and ML technologies into medical devices has revolutionized healthcare 
diagnostics and treatments. Medical device manufacturers are keen to maximize the advantages afforded 
by AI and ML by consolidating multiple applications onto a single platform. However, concurrent 
execution of several AI applications, each with its own visualization 
components, leads to 
unpredictable end-to-end latency, primarily due to GPU resource contentions. To mitigate this, 
manufacturers typically deploy separate workstations for distinct AI applications, thereby increasing 
financial, energy, and maintenance costs. This paper addresses these challenges within the context of 
NVIDIA's Holoscan platform, a real-time AI system for streaming sensor data and images. We propose a 
system design optimized for heterogeneous GPU workloads, encompassing both 
compute and 
graphics tasks. Our design leverages CUDA MPS for spatial partitioning of compute workloads and 
isolates compute and graphics processing onto separate GPUs. We demonstrate significant performance 
improvements across various end-to-end latency determinism metrics through empirical evaluation with 
real-world Holoscan medical device applications. For instance, the proposed design reduces 
maximum latency by 21--30\% and improves latency distribution flatness by 17--25\% for up to five 
concurrent endoscopy tool tracking AI applications, compared to a single-GPU baseline. Against
a default multi-GPU setup, our optimizations decrease maximum latency by 35\% for 
up to six concurrent applications by improving GPU utilization by 42\%. This paper provides clear 
design insights for AI applications in the edge-computing domain including medical systems, where 
performance predictability of concurrent and heterogeneous GPU workloads is a critical requirement.
\end{abstract}


\vspace{-4pt}
\section{Introduction}
\vspace{-2pt}
\label{sec:intro}
The integration of Artificial Intelligence (AI) and Machine Learning (ML) 
technologies into the medical devices represents a paradigm shift in the
healthcare and diagnostics industry. AI and ML applications have gained great 
traction in recent years by enhancing medical procedures, ranging from disease
diagnostics to surgical interventions. They empower automated real-time 
monitoring and provide valuable insights to medical professionals and 
physicians, leading to early diagnosis of diseases and reduced patient 
recovery times, among other benefits~\cite{cherubini2023review,ali2022we,catto2022effect}. These 
medical AI applications need both accelerated AI workload processing and optimized graphical 
rendering, enabled by modern GPUs, which has given rise to innovative 
platforms like the \nvidiaholoscan{}.

\holoscan{} is \nvidia{}'s scalable edge-computing platform for AI-enabled sensor processing. It 
empowers medical device manufacturers with the ability to create real-time
pipelines for AI-based analysis and visualization of streaming data and 
medical images~\cite{holoscan_sdk}. \holoscan{} 
consists of optimized libraries for I/O, data processing, inference and graphical rendering on the 
\nvidia{} GPUs. It provides a modular and intuitive data-flow programming model and is compatible with 
both ARM- and x86-based hardware equipped with GPUs, offering medical device vendors the freedom to 
select their architecture.

As the field of AI continues its rapid evolution, developers aspire to harness the 
potential of \holoscansdk{} for integrating \emph{multiple} AI
applications with visualization capabilities into their systems. This is especially true for 
edge-computing domains like medical devices, as compute workloads (enabled by CUDA) in this domain 
cannot leverage the full potential of massive parallelism offered by today's 
GPUs~\cite{zhao2018deepthings,dhakal2020gslice}.
Therefore, a manufacturer may seek to concurrently run multiple \holoscan{} AI
applications~\cite{intuitive_ion,olympus_eus}, aiming to enhance the efficiency of medical procedures. 
Real-world constraints such as space limitations, power consumption, financial cost 
considerations, maintenance, and regulatory requirements also necessitate utilizing as 
minimized set of hardware as possible, like a single workstation with GPU(s). 

\begin{figure}[H]
	\centering
	\vspace{-10pt}
	\includegraphics[width=0.8\columnwidth]{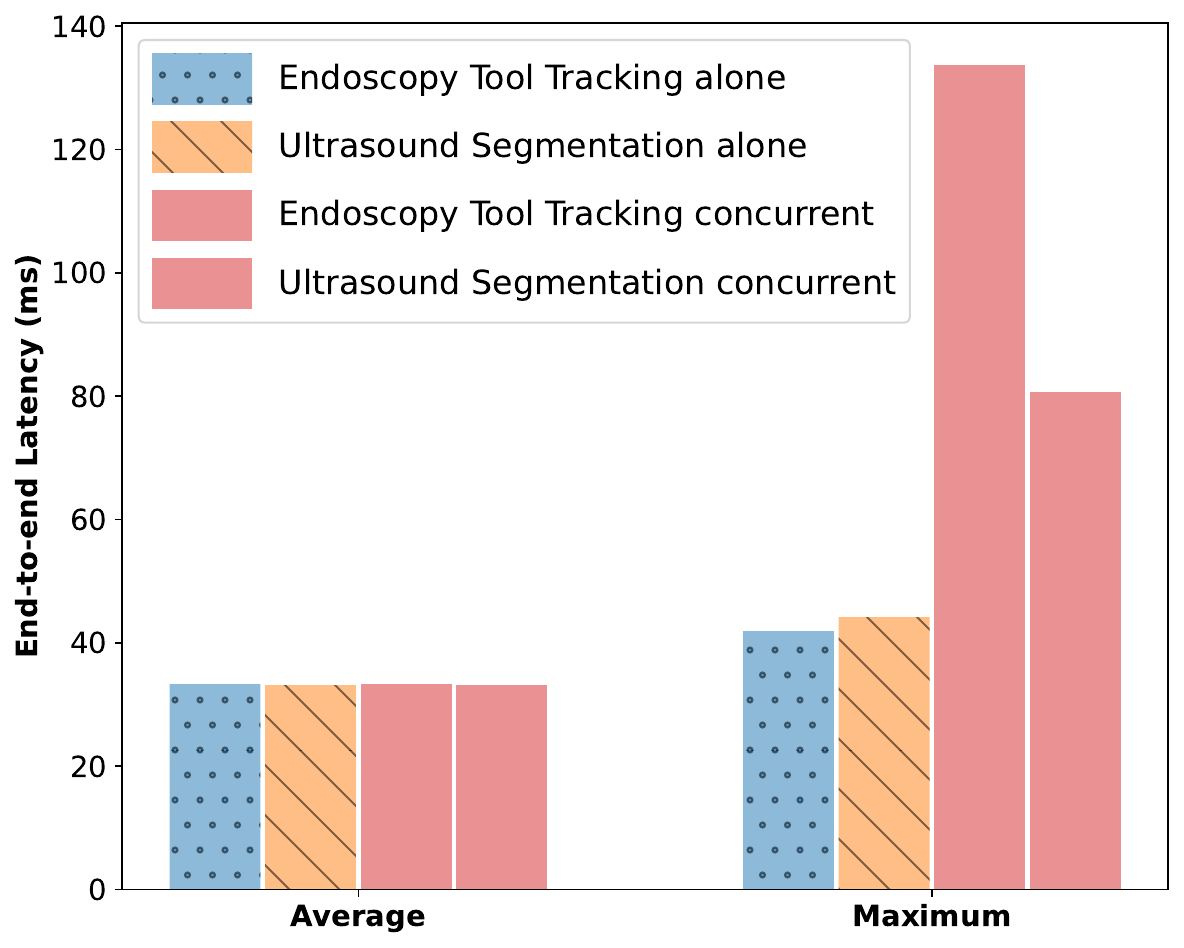}
	\vspace{-2pt}
	\caption{Average and Maximum End-to-end Latency on an x86 Workstation with an RTX A4000 GPU}
	\label{fig:endoultra}
	\vspace{-10pt}
\end{figure}

These constraints, however, 
often pose challenges to performance predictability~\cite{steimers2022sources}, since heterogeneous 
nature of the simultaneous AI-based compute and graphical rendering workloads create resource 
contention on a GPU~\cite{nickolls2009graphics}.
For example, Figure~\ref{fig:endoultra} shows the average and maximum end-to-end (E2E) latency when 
running endoscopy tool tracking and ultrasound segmentation \holoscan{} applications alone and 
concurrently on an x86 workstation with an RTX A4000 GPU. Simultaneous execution 
of multiple AI and visualization workloads leads to increased maximum 
latency. To avoid this issue, 
device manufacturers employ distinct workstations for separate AI applications, increasing the 
economic burden on themselves, hospitals and ultimately patients.

This paper addresses the critical challenge of deterministic end-to-end 
latency in medical AI systems with concurrent and heterogeneous GPU 
workloads. The primary contributions of the paper are 
outlined as follows:
\begin{itemize}[leftmargin=!,labelindent=0.5pt,align=left]
	\item We propose a novel design approach that combines CUDA MPS (\textbf{M}ulti-\textbf{p}rocess 
	\textbf{S}ervice)~\cite{cuda-mps} for spatial 
	partitioning between compute workloads and a load-balancing technique to isolate compute (CUDA 
	kernel) and graphics tasks onto distinct GPUs. Additionally, we use an 
	admission control policy to prevent 
	SM-oversubscription by concurrent compute tasks. Our pragmatic design is 
	straightforward to implement 
	and minimizes heavy context-switch overheads~\cite{kubisch2015life}, mitigating the resource 
	contention within a GPU for medical AI workloads.
	
	\item Empirical evaluation using a set of E2E latency determinism metrics 
	reveals substantial 
	performance improvement with our design: maximum latency, standard 
	deviation, latency distribution tail, and flatness are reduced by 
	21--30\%, 17--24\%, 21--47\%, and 17--25\%, 
	respectively, for up to five concurrent endoscopy tool tracking AI 
	applications, against a traditionally used 	single-GPU workstation. 
	Compared to a baseline multi-GPU system, our design decreases the maximum 
	latency by 35\% for up to six concurrent applications by improving GPU 
	utilization by 42\%.
\end{itemize}

Our design and results hold the promise of facilitating more
AI deployments in medical systems with more safety and predictability, while 
also providing design suggestions for other embedded and edge-computing 
application pipelines seeking to leverage concurrent 
and heterogeneous GPU workloads.

Next section provides an
overview of the \holoscansdk{}, followed by its usage in medical devices in Section~\ref{sec:medical}. 
Our system design approach for predictable performance is detailed in Section~\ref{sec:deterministic}. 
Section~\ref{sec:experimental_setup} describes our experimental setup for performance evaluation, 
followed by a comprehensive analysis of the experimental results using 
real-world, industry-grade \holoscan{} medical applications in Section~\ref{sec:result}. Finally, we 
discuss the related work in Section~\ref{sec:related} and conclude the paper in
Section~\ref{sec:conclusion}.

\vspace{-2pt}
\section{\holoscansdk{}}
\label{sec:holoscan}

\nvidia{} \holoscan{} SDK is an open-source AI sensor processing platform designed for low-latency 
real-time 
steaming data and images~\cite{holoscan_sdk}. It includes optimized libraries for data processing and 
AI, as well as core 
software services for various applications ranging from embedded systems to edge computing
platforms. Initially designed for medical devices, the SDK has evolved to be domain-agnostic, capable 
of serving multiple sectors like High-performance Computing at the Edge, Industrial Automation and 
Space Computing. It offers both C++ and Python APIs streamlining application prototyping and 
production deployment. It also includes built-in capabilities for functionalities such as I/O, ML 
inference, processing, and visualization, optimized for \nvidia{} GPUs.

A typical \holoscan{} application acquires and processes streaming data and images, and either 
controls an actuator or renders the processed output to a screen. An application consists of a number 
of \textit{fragments} where each fragment is a connected graph of \textit{operators}. Each fragment 
is assigned to a physical or virtual machine. For this paper, we only focus on single-fragment 
applications in this work. 

An operator is a core execution unit for a specific task. Operators interact with each other through 
\textit{Ports} which are data entry and exit points for an operator. An operator ingests streaming 
data at an input port and transmits data via an output port. The \holoscansdk{} provides a number of 
built-in operators, such as \textit{FormatConverterOp} for converting data format, 
\textit{InferenceOp} 
for AI inferences~\cite{holoscan_operator}. Application developers can also write their own operators 
by implementing initialization (\texttt{start()}), processing (\texttt{compute()}) and completion 
(\texttt{stop()}) functions of an operator. A single-fragment \holoscan{} application is created by 
connecting multiple operators in a graph to handle streaming AI workloads.

A \holoscan{} application is executed by topologically sorting the operators in a connected 
graph-fragment on a single thread.\footnote{Multithreaded execution in \holoscan{} is very recently 
enabled.} It ensures that the data processing of previous operators is completed before the next 
ones are executed. Each \holoscan{} operator has a set of \textit{Conditions} for scheduling 
decisions. Detailed discussions on these topics are not relevant to this work. Next, we describe two 
key components of the \holoscan{} 
SDK, both of which utilize the GPU - inference and visualization modules - 
before explaining the end-to-end latency measurements with the data flow 
tracking component.

\vspace{-2pt}
\subsection{Inference} 
\begin{figure}[h]
	\centering
	\footnotesize
	\vspace{-12pt}
	\includegraphics[width=1.0\columnwidth]{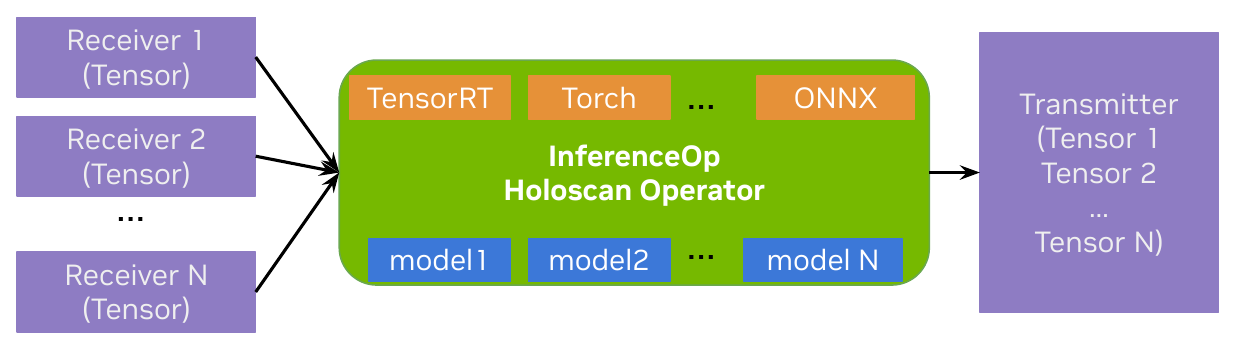}
	\vspace{-10pt}
	\caption{InferenceOp Operator in \holoscan{} SDK}
	\label{fig:inference_component}
	\vspace{-6pt}
\end{figure}
The inference module in the SDK provides APIs for designing inference-based 
AI applications. The \textit{InferenceOp} utilizes this module to load an AI model at runtime. It can 
also be configured with a range of parameters, including specifying inference backend, particular GPU 
device. It supports a variety of ML library backends, including TensorRT, Torch and ONNX runtime. The 
operator receives tensors into its input ports, performs inferences on the input tensors based on a 
given model for an input port, and finally emits output data as a processed tensor (see 
Figure~\ref{fig:inference_component}). Multiple inference workloads are concurrently launched in 
parallel CUDA streams on separate CPU threads.

\vspace{-2pt}
\subsection{Visualization} 
%
Holoviz is the visualization component in the \holoscan{} SDK that allows 
developers to seamlessly 
create and display advanced visual effects with minimal effort. It can combine real-time 
streams of frames in multiple layers including image stream, segmentation mask, geometry, text and GUI 
layers. It also supports depth map and volumetric 3d rendering, pertinent for robotic surgery 
and other advanced medical procedures. Holoviz leverages the Vulkan graphics 
API~\cite{sellers2016vulkan} to optimize rendering performance and adopts the immediate mode design 
pattern~\cite{dear_imgui} to accelerate the rapid visualization updates in medical devices. 
\textit{HolovizOp} operator can be used to connect and transfer 
tensors and video buffers from other operators for visualization. Since it 
utilizes the default CUDA stream 
for compute operations, resource contention between inference and other CUDA 
compute operations may occur in absence of any isolation between graphics and 
compute contexts.

\vspace{-4pt}
\subsection{Data Flow Tracking}
Inspired by the ideas of Information Flow Tracking~\cite{denning1976lattice}, \holoscan{} 
implements data flow tracking within its fragment-graph. With this feature, each incoming and outgoing 
message in an operator is timestamped. The end-to-end latency of individual messages is calculated 
using these fine-grained timestamps. End-to-end 
latency~\cite{sinha2022solution} is the time interval between a message's 
arrival at a root operator of the fragment-graph via camera or other sensory 
input and its departure from a leaf operator for either rendering or 
actuation. We evaluate \holoscan{} 
applications through a set of performance determinism metrics based on the 
observed end-to-end latencies using the data flow tracking component.

\vspace{-2pt}
\section{Medical Device Application Use-cases}
\label{sec:medical}

\holoscansdk{} is useful for different edge-computing domains, including industrial automation 
and space computing, but medical AI is one of its primary target areas. There are several 
available \holoscan{} medical AI applications~\cite{holohub}. This section describes two example
applications to provide a high-level overview of their functionalities.

\vspace{-4pt}
\subsection{Endoscopy Tool Tracking Application} 
\begin{figure}[b!]
	\centering
	\vspace{-12pt}
	\includegraphics[width=\columnwidth]{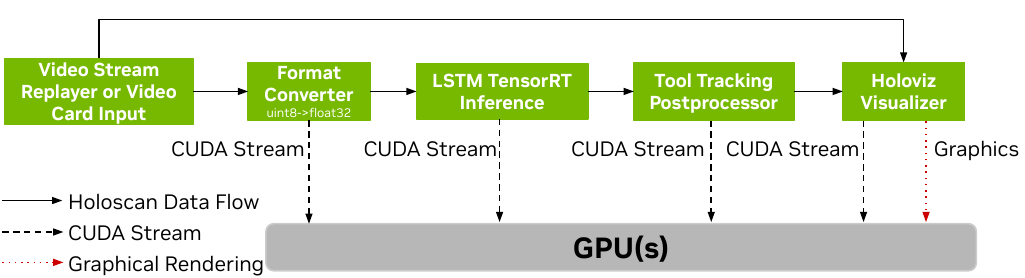}
	\vspace{-6pt}
	\caption{Endoscopy Tool Tracking Application}
	\label{fig:endoscopy_tool_tracking}
\end{figure}
The Endoscopy Tool Tracking application is designed to detect and annotate endoscopic instruments 
within a live stream of video frames. It ingests a video stream input either via a \textit{Replayer} 
or a custom video card~\cite{kona_xm} operator. The video stream is converted into a tensor and handed 
over to a \textit{FormatConverter} operator which changes the datatype of the tensor 
elements as part of pre-processing. The tensor is then fed into an \textit{LSTMTensortRTInference} 
operator for inference via TensorRT~\cite{tensorrt}. 
\textit{LSTMTensortRTInference} loads a pre-trained inference 
model on a GPU at runtime. The output tensor containing the inference result is then processed by a 
\textit{ToolTrackingPostProcessorOp} operator for endoscopic tool segmentation. Finally, a 
\textit{Holoviz} operator overlays the segmented tools and their text labels onto the original input 
video frame for visualization. Figure~\ref{fig:endoscopy_tool_tracking} summarizes the application 
data flow and its interaction with GPU(s).


The application is implemented in both C++ and Python. 
A snippet of the C++ code is provided below. It creates an \texttt{App} class 
derived from \texttt{Application} class. The \texttt{compose()} member 
function instantiates the required \holoscan{} operators. The 
operator are configured with the \texttt{from\_config} function that reads a YAML configuration file. 
CUDA stream for an operator is allocated with
\texttt{CudaStreamPool} class which manages data-transfer and compute operations on the GPU(s). A CUDA 
stream is initialized once for \texttt{FormatConverterOp} and reused by subsequent operators in the 
pipeline.

\vspace{-2pt}
\begin{lstlisting}[language=C++]
class App : public holoscan::Application {
public:
...
void compose() override {
...
auto source = 
  make_operator<ops::VideoStreamReplayerOp>(
  "replayer", from_config("replayer"));

auto format_converter = 
  make_operator<ops::FormatConverterOp>(
  "format_converter",...,
  Arg("cuda_stream_pool") = cuda_stream_pool);

auto lstm_inferer = 
  make_operator<ops::LSTMTensorRTInferenceOp>(
  "lstm_inferer",...);

auto tool_tracking_postprocessor = 
  make_operator<ops::ToolTrackingPostprocessorOp>(
  "tool_tracking_postprocessor");

auto visualizer = 
  make_operator<ops::HolovizOp>("holoviz",...);
...
}
}
\end{lstlisting} 
\vspace{-2pt}

After initialization, operators are connected in \texttt{compose} to define 
the data flow using the \texttt{add\_flow} function. The 
first two parameters specify source and destination operators, while the third 
parameter defines the port map connecting them.

\vspace{-2pt}
\begin{lstlisting}[language=C++]
// Flow definition
add_flow(source, format_converter, 
  {{"output", "source_video"}});
add_flow(format_converter, lstm_inferer);
add_flow(lstm_inferer, 
  tool_tracking_postprocessor,{{"tensor", "in"}});
add_flow(tool_tracking_postprocessor, visualizer, 
  {{"out", "receivers"}});
\end{lstlisting}
\vspace{-4pt}

\subsection{Multi AI Ultrasound Application} 
\label{sec:multiai_ultrasound}
The Multi-AI Ultrasound application utilizes three machine learning models from iCardio~\cite{icardio} 
concurrently for real-time cardiovascular ultrasound diagnostics. It uses four instances of 
\textit{FormatConverterOp}, one for each model and an additional one for visualization. The 
\textit{InferenceOp} facilitates parallel inferences across multiple CUDA 
streams, each spawned from 
individual CPU threads. The inference models perform anatomical heart component measurements and 
classification for various cardiac views and aortic stenosis anomalies.
\textit{InferenceProcessorOp} executes post-processing operations, while \textit{HolovizOp} renders 
the aggregated overlaid layers along with the original video frames. Unlike the 
endoscopy example, this application consists of multiple paths in its graph of operators.

\vspace{-2pt}
\section{Deterministic Design Approach}
\label{sec:deterministic}
In this section, we describe our design to enhance 
performance determinism within the \nvidiaholoscan{} platform when applied to 
medical AI applications. These techniques primarily focus on mitigating 
resource contention arising from concurrent and heterogeneous GPU workloads. Our design
spatially isolates GPU workloads including graphics on dedicated SMs.
Our design and optimization techniques, although studied in the context of medical AI, are generally 
applicable to heterogeneous GPU workloads.

\vspace{-2pt}
\subsection{CUDA MPS Partition}
CUDA MPS~\cite{cuda-mps} is 
NVIDIA's technology designed for concurrent execution of multiple GPU-accelerated 
applications on GPUs. It leverages the Hyper-Q 
capability~\cite{bradley2012hyper} present in NVIDIA GPUs, enabling the 
concurrent processing of multiple CUDA kernels on the same GPU. This 
functionality helps in embedded and edge computing applications like medical 
devices software, where GPU workloads 
predominantly involve inference and lightweight data processing tasks and do 
not saturate the computational capabilities of today's GPUs like the RTX 
A4000 and A6000.

Traditionally, a CUDA program that wants to run a GPU workload begins by 
creating a CUDA context or using the default primary CUDA context, 
encapsulating the essential hardware 
resources. Multiple GPU workloads can be organized into CUDA streams in a single CUDA context. 
Furthermore, multiple processes can create separate CUDA contexts but 
depend on the GPU hardware scheduler for time-sliced GPU workload execution. The time-multiplexed 
execution of CUDA streams and contexts on the GPU and dependence on the GPU hardware scheduler for SM 
allocation may increase throughput at the cost of lower performance 
determinism.

CUDA MPS aims to better utilize the GPU for concurrently running GPU workloads with a client-server 
model. An MPS server is a single process that manages the GPU resources and coordinates the execution 
of multiple MPS clients' CUDA tasks. The CUDA kernel launches are intercepted by the MPS server to be 
launched on behalf of the client. This allows the MPS server to optimize 
GPU resource allocation and run concurrent CUDA kernels in separate CUDA contexts.

Before the NVIDIA Volta GPU architecture, the MPS server was responsible for 
managing hardware resources for a single CUDA context. All the client CUDA 
contexts from multiple processes would route their CUDA workloads through the MPS 
server. 
Since the Volta architecture, MPS client CUDA contexts directly  
manage most hardware resources. The MPS server continues to play a role in 
mediating the shared resources. 
The communication between MPS clients and the server remains 
transparent to applications, facilitated through the usual
CUDA API. Existing CUDA programs do not require any 
modification to utilize CUDA MPS features. 


We utilize CUDA MPS functionalities to establish isolated partitions resembling GPU
\textit{sandboxes} consisting of exclusive SMs and device memory.
To ensure deterministic application performance, a CUDA program requires, at 
least, exclusive access to its required number of SMs and memory, so that concurrent CUDA tasks do not 
interfere with or hamper each other's performance. CUDA MPS partitioning enables it.

\subsubsection{MPS Partition Creation}
By default, all SMs in a GPU are available to any MPS client. This may lead to potential latency 
variability as concurrent CUDA kernels may contend for the same set of SMs. 
CUDA MPS provides a mechanism for constraining 
the allocation of GPU resources to each
MPS client process, with the \texttt{CUDA\_MPS\_ACTIVE\_THREAD\_PERCENTAGE} and 
\texttt{CUDA\_MPS\_PINNED\_DEVICE\_MEM\_LIMIT} environment variables to,
respectively, limit the number of SMs and device memory.
We employ this mechanism to create isolated GPU partitions for each MPS client. The required number of 
SMs and device memory are determined through offline 
profiling~\cite{wilhelm2008worst} of a \holoscan{} application.
Future work will consider cache, memory bus, and other microarchitecture partitioning.

\subsubsection{Admission Control}
To avoid GPU overload and contention possibility caused by an excessive number of MPS clients, we 
implement an admission control that denies launching CUDA workload for an MPS client if its SM 
requirements surpass the currently available SMs. As CUDA MPS does not 
support strong SM affinity~\cite{bakita2023hardware}, our 
approach prevents resource sharing, reducing further interference between applications.

\vspace{-2pt}
\subsection{Hardware Isolation Between Compute and Graphics}
\begin{table}[b]
	\centering
	\renewcommand{\arraystretch}{1.2} 
	\footnotesize
	\vspace{-12pt}
	\caption{Feature Comparison between MIG, vGPU and CUDA MPS for Deterministic Performance against 
		Heterogeneous (Compute and Graphics) GPU Workloads}
	\label{tab:mig_vgpu_mps_comparison}
	\vspace{-2pt}
	\begin{tabularx}{\columnwidth}{|X|l|c|c|}\hline
		\textbf{Feature} & \textbf{MIG} & \textbf{vGPU} & \textbf{CUDA MPS}\\\hline\hline
		\textbf{Dedicated SM Allocation} \newline (No Time-slicing) &  
		\multicolumn{1}{c|}{\textbf{\color{white}\cellcolor{nvidiacolor}Yes}} & \cellcolor{lightred} 
		No & 
		\multicolumn{1}{c|}{\textbf{\color{white}\cellcolor{nvidiacolor}Yes}} \\\hline
		\textbf{Custom SM Allocation} & \cellcolor{lightred} No & \cellcolor{gray} - & 
		\multicolumn{1}{c|}{\textbf{\color{white}\cellcolor{nvidiacolor}Yes}} \\\hline
		\textbf{Graphics Support} & \cellcolor{lightred} No & 
		\multicolumn{1}{c|}{\textbf{\color{white}\cellcolor{nvidiacolor}Yes}} & 
		\cellcolor{lightred} No \\\hline
		\textbf{Supported in RTX A4000 or A6000}\newline(or other ProViz GPUs) & 
		\cellcolor{lightred} No & 
		\multicolumn{1}{c|}{\textbf{\color{white}\cellcolor{nvidiacolor}Yes}} & 
		\multicolumn{1}{c|}{\textbf{\color{white}\cellcolor{nvidiacolor}Yes}}
		\\\hline
	\end{tabularx}
\end{table}
GPU workloads in medical AI systems are often heterogeneous, comprising both
graphics rendering and compute tasks. However, multiplexing these heterogeneous workloads
on a single GPU can lead to nondeterministic performance. The NVIDIA GPUs incorporate an 
optimized and coherently structured graphics pipeline that 
cannot be externally preempted or controlled~\cite{kubisch2015life}. As medical equipment 
applications consider both graphics and compute equally important, it is critical to ensure 
timing-predictable rendering and compute task output.
Technologies like Multi-instance GPU (MIG), virtual GPU 
(vGPU)~\cite{mig_vgpu}, and CUDA MPS are limited in supporting predictable performance 
against heterogeneous GPU workloads, and they are summarized in 
Table~\ref{tab:mig_vgpu_mps_comparison}. Although MIG provides strong hardware isolation, 
it does not support graphics. vGPU in time-slicing mode cannot reserve exclusive SMs. We are using 
CUDA MPS partitioning with a maximum cap on the number of dedicated SMs, but it does not handle 
graphics.

It is clear that existing solutions do not isolate graphics and compute on dedicated SMs. To address 
this, we divide 
the graphics and compute workloads to separate physical GPUs. Although this introduces the overhead 
of transferring data from the compute GPU memory to the graphics GPU memory, 
\nvidia{}'s Unified Virtual Addressing~\cite{schroeder2011peer} is 
leveraged to initiate zero-copy transfer with minimal overhead. Additionally, the data-transfer cost 
amortizes with more workload because of overlapped data-transfer and 
compute~\cite{cuda_overlap,schroeder2011peer}. 
The experiments demonstrate that isolating two types of GPU workloads to 
distinct GPUs is an effective mitigation of resource contention in this case. 
Generally, using a more compute-capacity GPU 
for compute and a lesser one for graphics is advised since compute workloads are more 
SM-intensive. A compute GPU is used for executing the
CUDA kernels including data pre- and post-processing, AI inferences. The 
graphics GPU is mainly used for rendering, which leverages Vulkan APIs via 
\textit{Holoviz}.

\vspace{-4pt}
\subsection{CPU Affinity}
Medical device systems typically pre-determine the number and types of applications for deployment.
Such foresight allows for pinning each application to specific CPU cores to reduce Linux 
scheduling overhead, as advised for safety-critical systems~\cite{cerrolaza2020multi}.
We utilize Linux CPU affinity APIs (\texttt{sched\_setaffinity}), and utilities (\texttt{taskset}, 
\texttt{cset}, \texttt{isolcpus}) for this purpose.

\begin{figure}[b]
	\centering
	\vspace{-12pt}
	\includegraphics[width=0.9\columnwidth]{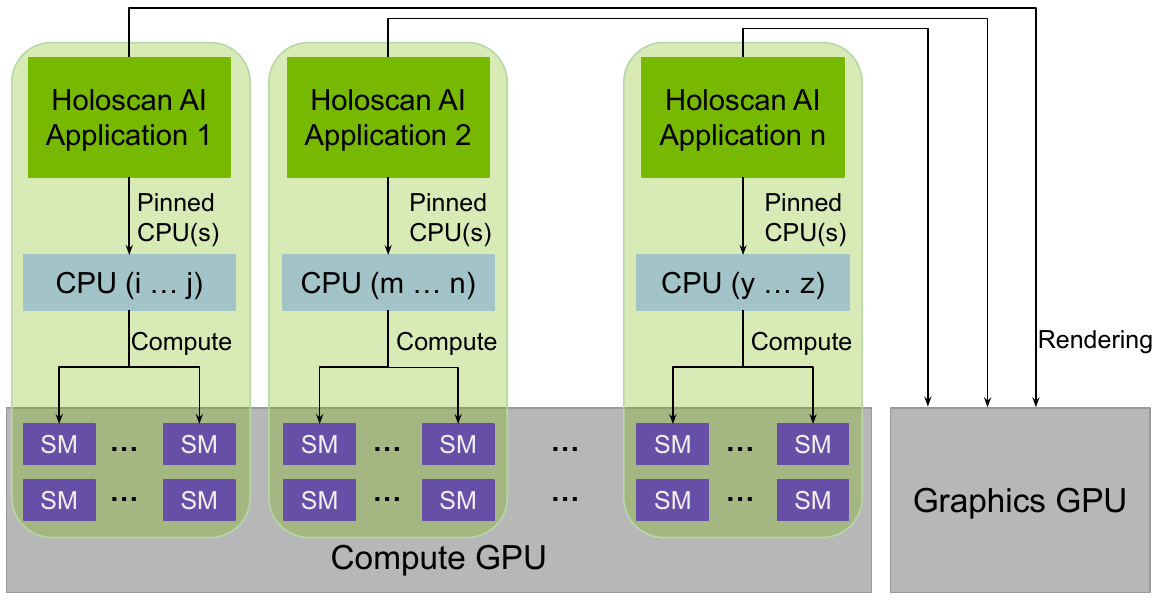}
	\vspace{-2pt}
	\caption{System Design for Concurrent AI Applications}
	\label{fig:runtime_deterministic_arch}
\end{figure}

\vspace{-4pt}
\subsection{Design Summary}
Figure~\ref{fig:runtime_deterministic_arch} summarizes our system design proposal for
concurrent \holoscan{} applications. Every \holoscan{} 
application is launched as a separate process, that could be pinned to a set of CPU cores. Their GPU 
workloads are divided into distinct GPUs based on the workload type. Their compute workloads
are launched to a number of dedicated SMs on a compute GPU, enabled by CUDA MPS partitioning, while 
rendering is directed to a graphics GPU. The resultant workload partition is elaborated as an 
addendum in Appendix (Figure~\ref{fig:strategy_overview}).

\vspace{-4pt}
\subsection{Scope and Limitations}
We discuss our solutions in the context of NVIDIA GPUs and the \holoscansdk{} for medical AI devices 
because of their widespread adoption, accessibility, and long-term 
software support. However, these techniques are also applicable to other 
GPUs~\cite{otterness2021exploring} and heterogeneous workloads. Furthermore, this work mainly focuses 
on compute resource (SM) isolation. Future work will explore mitigation for cache, memory 
bus, and other shared resource contention~\cite{ausavarungnirun2018mask}.

\vspace{-2pt}
\section{Experimental Setup}
\label{sec:experimental_setup}
\holoscan{} officially supports ARM-based NVIDIA IGX Orin devkit with RTX 
A6000 GPU, and x86 workstations with selected discrete GPUs. Typically, 
medical device manufacturers use a high-performance x86 workstation with a 
discrete GPU. For this paper, we used an Intel i7-7920X 2.90 GHz CPU, 
equipped with GPUs: RTX A4000 (48 SMs, 6144 CUDA cores, 19.2 
single-precision (SP)
TFLOPS) and A6000 (84 SMs, 10752 CUDA cores, 38.7 SP TFLOPS). 
Although the emerging IGX Orin platform has gained significant traction for 
medical AI~\cite{active_surgical,medtronic_igx}, it does not 
support CUDA MPS for now~\cite{cuda-mps}. Future work will consider it with 
inter-operable iGPU and dGPU for multi-GPU and MPS configurations.

\vspace{-6pt}
\subsection{End-to-end Latency Determinism Metrics}
\label{sec:e2e_latency_metrics}
E2E latency, a well-known performance indicator in real-time
cyber-physical systems, is the time-interval between a message's arrival to and 
departure from an application. In \holoscan{}, a message may traverse through 
multiple paths in the graph of operators. We consider every path's E2E latency.

To gauge the overall performance quality of an 
application, we examine the average E2E latency, which also indicates the 
frame-rate for visualization. In contexts where safety, reliability, and 
certification concerns are paramount, the maximum E2E latency is a primary 
metric, used also for latency determinism. Although the industry, especially 
the medical devices sector, rarely uses any other metrics to quantify latency 
determinism~\cite{mori2018real,luo2019real}, we use three more key metrics 
to assess it.

\begin{enumerate}[leftmargin=!,labelindent=0pt,align=left]
	\item \textbf{Standard Deviation:} It measures the spread of
	latencies around the average value. A lower standard deviation indicates 
	concentrated latencies around the mean latency. However, it does not 
	provide information about the overall latency distribution and outliers like 
	the maximum latency.
	
	\item \textbf{Latency Distribution Tail:} It is the difference between the
	95th and 100th percentile 
	values of the observed E2E latencies. 
	It indicates how widely the outlier latencies
	are distributed at the higher end of the distribution. A smaller 
	tail means that the outlier latencies are less stretched out near the 
	maximum latency, indicating greater predictability.
	
	\item \textbf{Latency Distribution Flatness:} It is defined as the difference 
	between the 10th and 90th percentile values of the observed end-to-end 
	latencies. A smaller flatness indicates more concentrated and 
	deterministic most-observed latencies.
\end{enumerate}

\vspace{-2pt}
\subsection{Applications}
We carried out most of our experiments with the Endoscopy Tool Tracking 
\holoscan{} application~\cite{holohub}. It is representative of a typical 
medical device AI application.
For experiments, we use a \holoscan{} operator that replays a saved video on the 
disk. However, \nvidiaholoscan{} also supports live video 
feed from high-fidelity cameras directly to the GPU memory 
through DMA~\cite{kona_xm,deltacast} which minimizes I/O overhead and facilitates 
faster data-processing in a GPU. Accelerated I/O solutions such as RDMA
improves the performance further, but these topics are out of scope of this 
paper. 

We experimented with two additional applications. The Ultrasound 
Segmentation pipeline~\cite{holohub} is similar to the Endoscopy 
application and 
performs automatic segmentation of the spine from a trained AI model for 
scoliosis visualization and measurement. The Multi-AI Ultrasound 
application~\cite{holohub}, 
described in Section~\ref{sec:multiai_ultrasound}, is 
more complicated than the others as it launches three AI inferences in 
parallel CUDA streams. 

For most experiments, we run the same applications from different processes 
to emulate a scenario of multiple Holoscan applications running on the same 
system. We experiment with up to five concurrent applications, sufficient for the 
needs of medical AI devices on a single platform.

\vspace{-2pt}
\section{Result Analysis}
\label{sec:result}
\begin{figure*}[!hb]
	\centering
	\vspace{-16pt}
	\begin{subfigure}[b]{0.32\textwidth}
		\centering
		\includegraphics[width=0.95\linewidth]{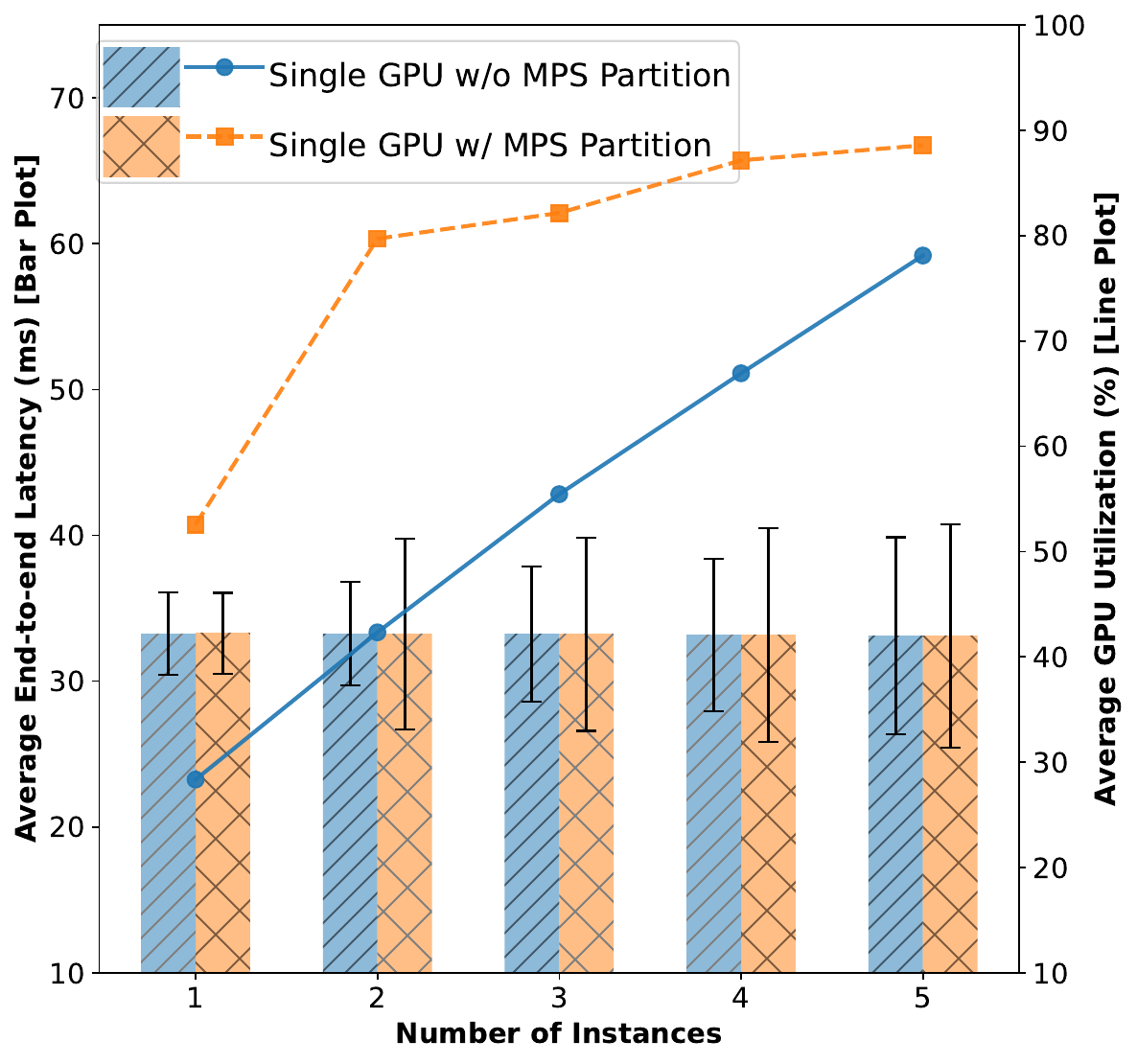}
		\caption{Average Latency and GPU Utilization}
		\label{fig:endoscopy_avg_x86_mps}
	\end{subfigure}
	\hfill
	\begin{subfigure}[b]{0.32\textwidth}
		\centering
		\includegraphics[width=\linewidth]{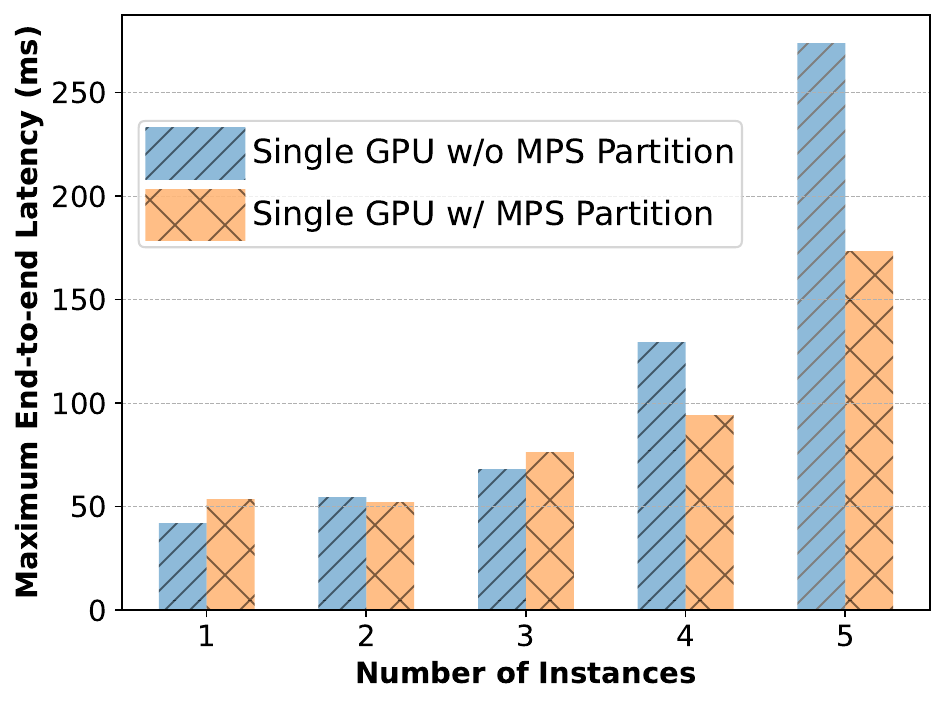}
		\caption{Maximum E2E Latency}
		\label{fig:endoscopy_max_x86_mps}
		\vspace{12pt}
	\end{subfigure}
	\hfill
	\hfill
	\begin{subfigure}[b]{0.32\textwidth}
		\centering
		\includegraphics[width=\linewidth]{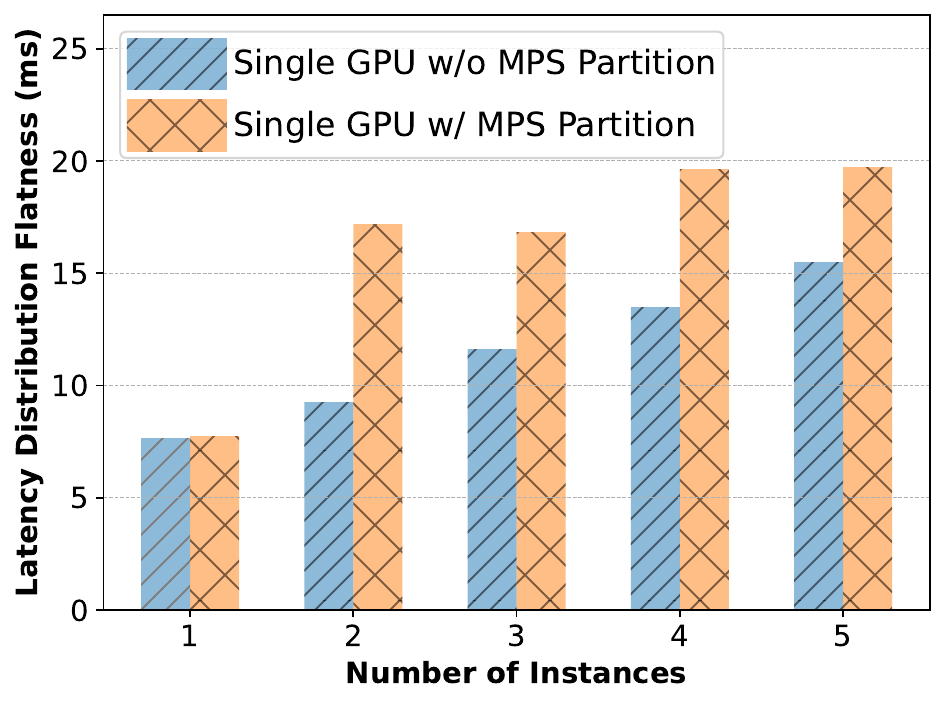}
		\caption{Latency Distribution Flatness}
		\label{fig:endoscopy_flatness_x86_mps}
		\vspace{12pt}
	\end{subfigure}
	\vspace{-4pt}
	\caption{Performance on Single GPU without vs. with CUDA MPS 
		Partitions for Endoscopy Tool Tracking}
	\label{fig:mps_gpu_endoscopy_x86}
\end{figure*}
\begin{figure*}[!t]
	\centering
	\footnotesize
	\begin{subfigure}[b]{0.32\textwidth}
		\centering
		\includegraphics[width=\linewidth]{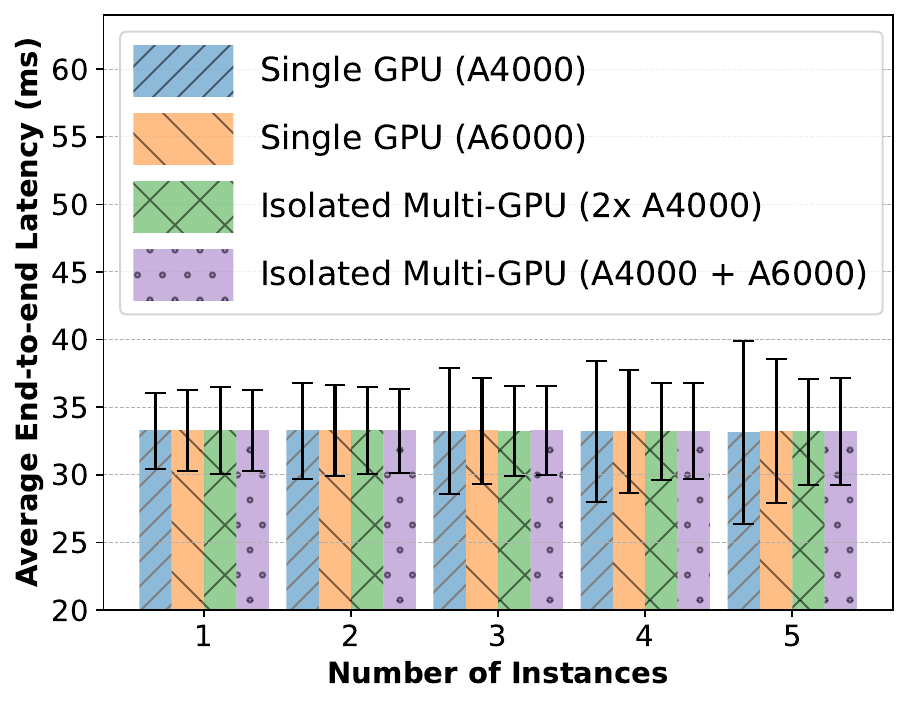}
		\caption{Average E2E Latency}
		\label{fig:endoscopy_avg_x86_multi}
	\end{subfigure}
	\hfill
	\begin{subfigure}[b]{0.32\textwidth}
		\centering
		\includegraphics[width=\linewidth]{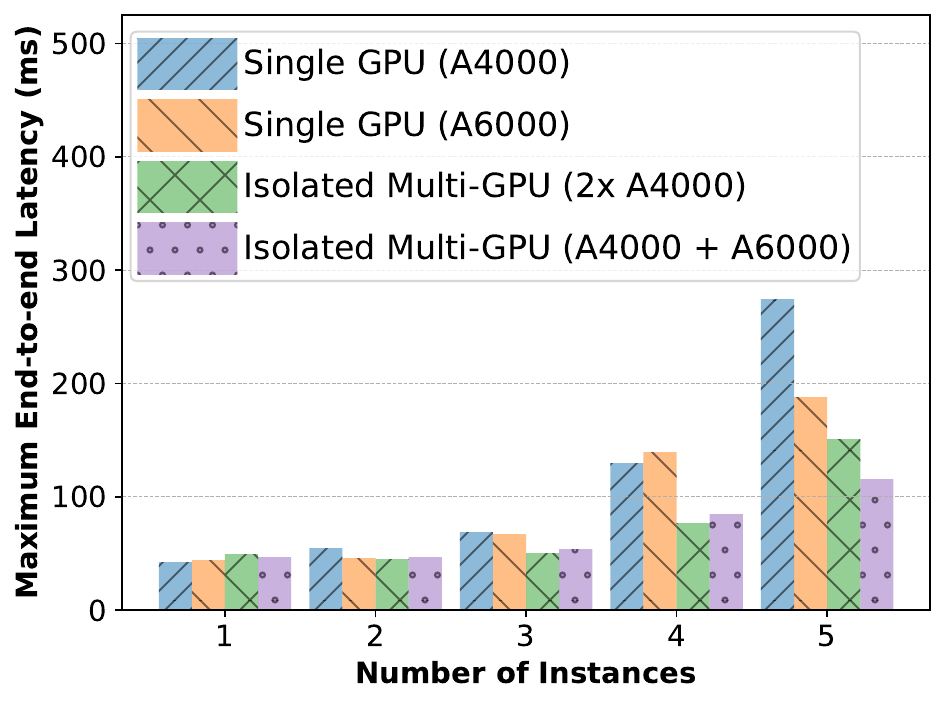}
		\caption{Maximum E2E Latency}
		\label{fig:isolated_gpu_endoscopy_x86_max}
	\end{subfigure}
	\hfill
	\begin{subfigure}[b]{0.32\textwidth}
		\centering
		\includegraphics[width=\linewidth]{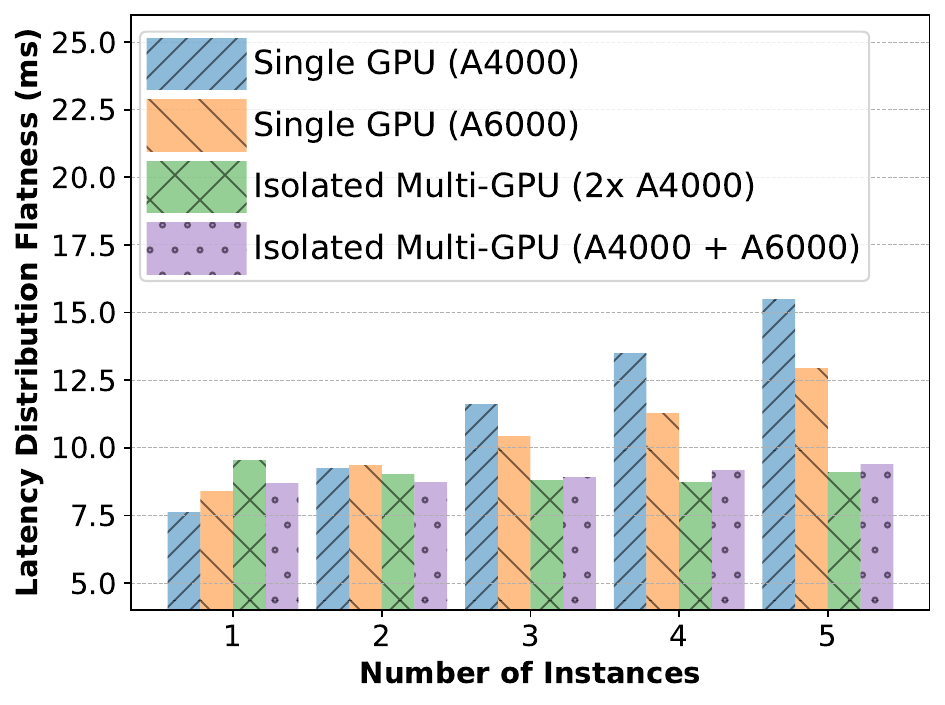}
		\caption{Latency Distribution Flatness}
		\label{fig:isolated_gpu_endoscopy_x86_flatness}
	\end{subfigure}
	\vspace{-4pt}
	\caption{Performance on Single GPU vs. Compute and Graphics Isolated on 
		Multi-GPUs for Endoscopy Tool Tracking}
	\label{fig:isolated_gpu_endoscopy_x86}
	\vspace{-16pt}
\end{figure*}
This section presents an experimental analysis of our system design. For 
every experiment, an application is executed for 
a period of 1000 messages or approximately 30 seconds, where the first and last 10 
messages are discarded in every experiment as warm-up and cool-down periods. Each 
experiment is repeated 10 times. We analyze our results using end-to-end (E2E) 
latency and GPU utilization metrics. Frame-rate for graphics rendering is 
inherently captured by the average E2E latency. Our analysis focuses 
on the latency distribution and dispersion, based on the metrics discussed in 
Section~\ref{sec:e2e_latency_metrics}.
The experimental artifacts will be made available later~\cite{holohub}.

\subsection{Performance Result with CUDA MPS Partitions}
\label{sec:cuda_mps_performance}
We first analyze the performance results with and without CUDA MPS partitioning on 
a single RTX A4000 GPU. For the experiments, we configured CUDA MPS partitions to 
limit 20\% of the GPU threads and 2GB of device pinned memory per client. 
These configurations were determined by profiling the application. 
Our experiments included up to five MPS clients as \holoscan{}
applications, ensuring that allocated threads and memory did not exceed the GPU's 
limits.

Figure~\ref{fig:endoscopy_avg_x86_mps} reveals that average latency (left Y-axis, 
bar plot) remains consistent 
for up to 5 concurrent endoscopy tool tracking instances, indicating a 
non-prohibitive amortized frame-rate.
However, the standard deviation, represented as error caps, rises with 
instance count. 
CUDA MPS yields similar or slightly elevated standard deviations 
compared to a non-MPS setup, as it does not handle contention between 
compute and graphics contexts. Figure~\ref{fig:endoscopy_avg_x86_mps} also captures
GPU utilization (right Y-axis, line plot) which is better with CUDA MPS 
than with the non-MPS setup, due to MPS' reservation on the number of 
GPU threads per client. More GPU utilization is adequately leveraged to 
improve the maximum E2E latency by up to a 37\%
(Figures~\ref{fig:endoscopy_max_x86_mps}), and latency distribution tail by up to 
45\% for five instances (Figure~\ref{fig:endoscopy_tail_x86_mps} in the 
Appendix). The benefits of 
MPS partitioning are amplified with more instances, as more 
resource contention deteriorates the performance predictability of a 
non-MPS single GPU setup. Figure~\ref{fig:endoscopy_flatness_x86_mps} 
shows no improvement in distribution flatness with CUDA MPS, 
due to software overhead and context-switching costs between graphics and 
compute contexts.

\vspace{-4pt}
\subsection{Performance Result with Compute and Graphics Isolation}
\label{sec:hardware_isolation_evaluation}
This section compares performance between two setups: 
one where compute and graphics workloads are segregated onto two distinct GPUs (2x 
A4000, and A4000+A6000), and another where all the tasks are handled by a single 
GPU (A4000 or A6000), both on a single x86 workstation. 
Figure~\ref{fig:endoscopy_avg_x86_multi} shows that the average E2E
latency stays consistent across single and multi-GPU configurations, enabled 
by average frame-rate of the source video.
However, the standard deviation increases with more 
instances in single-GPU setups but remains 
stable in workload-isolated multi-GPU arrangements, implying reduced 
context-switch costs and greater predictability.

Figure~\ref{fig:isolated_gpu_endoscopy_x86_max} demonstrates 16--24\% reduction 
in maximum E2E latency with our design of isolated multi-GPU configurations. 
Specifically, \textit{the dual A4000 configuration, using our workload 
partitioning design, achieves 16\% lower maximum latency than a single A6000 
setup, offering 7\% potential energy saving and over 50\% cost reduction}, based 
on specifications and prices listed on CDW and Amazon.com, and making medical AI 
applications more affordable and sustainable.

Although a 10\% latency penalty is incurred for 1 instance in the 
multi-GPU setups due to data-transfer overheads between the GPUs, this cost 
is offset with more instances~\cite{cuda_overlap,schroeder2011peer}. 
Moreover, latency distribution tail 
(Figure~\ref{fig:isolated_gpu_endoscopy_x86_tail} in 
Appendix) and flatness in Figure~\ref{fig:isolated_gpu_endoscopy_x86_flatness} are 
improved by 30\% and 
17\%, respectively, in multi-GPU setups. The result demonstrates that workload 
isolation in multi-GPUs overcomes the limitations of MPS partitioning by 
segregating the graphics and compute contexts. 
Section~\ref{sec:increased_gpu_eval} validates that these performance gains indeed 
stem from workload isolation, not merely from increased compute capacity with 
multiple GPUs.

\vspace{-2pt}
\subsection{Performance Result with All Optimizations}
\begin{figure*}[b!]
	\centering
	\vspace{-12pt}
	\begin{subfigure}[b]{0.32\textwidth}
		\includegraphics[width=\linewidth]{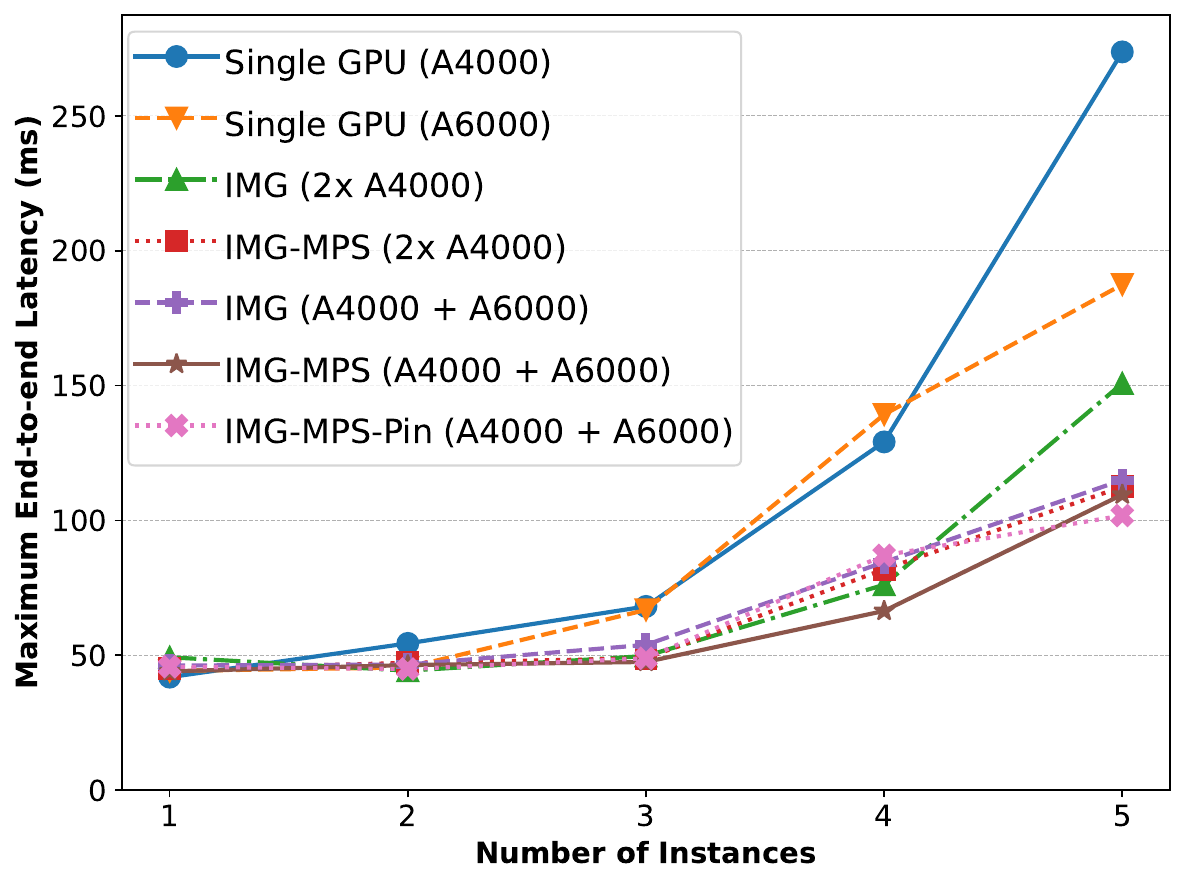}
		\vspace{-10pt}
		\caption{Maximum End-to-end Latency}
		\label{fig:endoscopy_max_x86_all}
	\end{subfigure}
	\hfill
	\begin{subfigure}[b]{0.32\textwidth}
		\includegraphics[width=\linewidth]{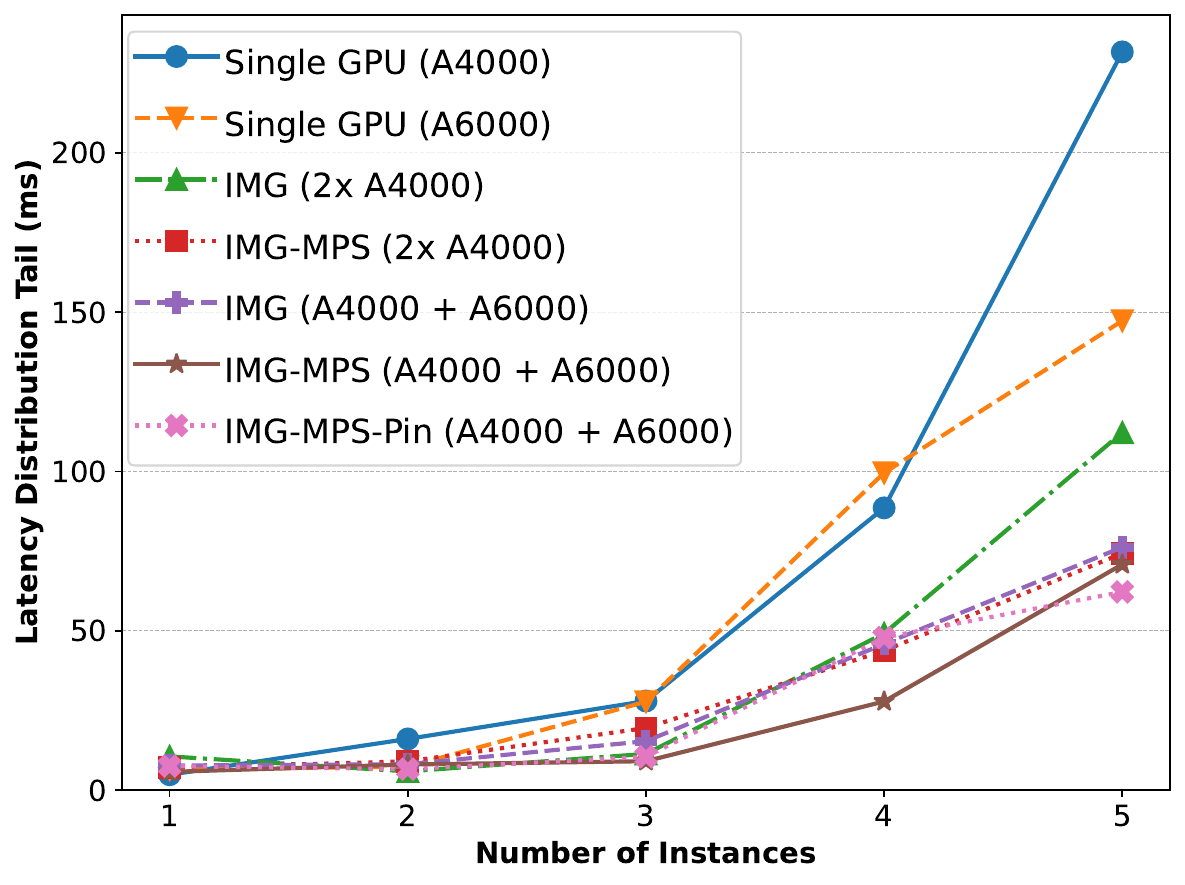}
		\vspace{-10pt}
		\caption{Latency Distribution Tail}
		\label{fig:endoscopy_tail_x86_all}
	\end{subfigure}
	\hfill
	\begin{subfigure}[b]{0.32\textwidth}
		\includegraphics[width=\linewidth]{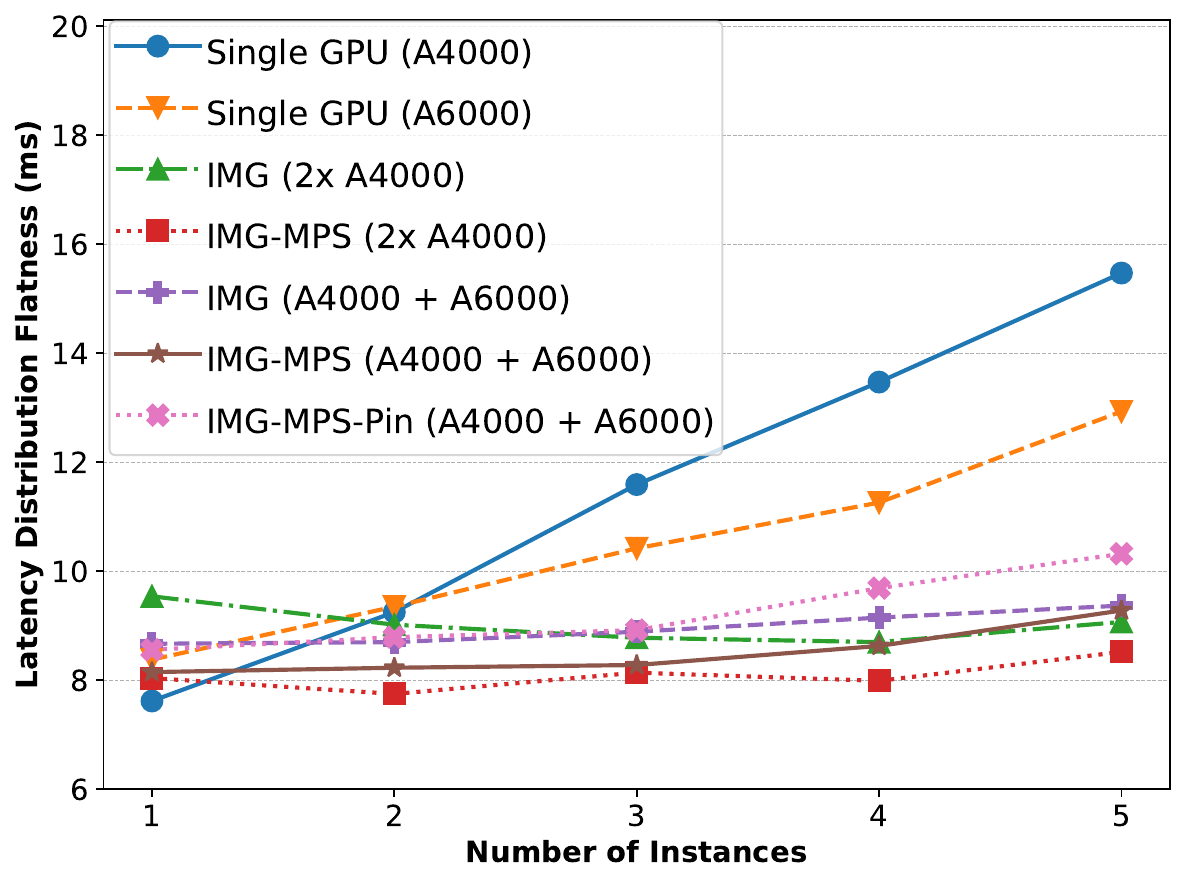}
		\vspace{-10pt}
		\caption{Latency Distribution Flatness}
		\label{fig:endoscopy_flatness_x86_all}
	\end{subfigure}
	\vspace{-4pt}
	\caption{Performance Comparison between All Optimizations for 
		Endoscopy Tool Tracking Application}
	\label{fig:endoscopy_x86_all}
\end{figure*}
In the subsequent experiments, we evaluate all configurations from our 
proposed design approaches:
\begin{enumerate}[leftmargin=*]
	\item \textbf{Single GPU}: Both compute and graphics workloads are executed on 
	a single GPU. This is the baseline, as manufacturers typically use such 
	configurations.
	
	
	\item \textbf{Isolated Multi-GPU (IMG)}: Compute and graphics tasks are 
	isolated onto distinct physical GPUs.
	
	\item \textbf{Isolated Multi-GPU + MPS Partition (IMG-MPS)}: Similar to 
	IMG but	MPS partitioning applied on top of it.
	
	\item \textbf{Isolated Multi-GPU + MPS Partition + CPU Pinning (IMG-MPS-Pin)}: 
	Extends IMG-MPS by pinning each application instance to a specific CPU core.
\end{enumerate}
\vspace{-2pt}
Average latency is similar across different configurations, like before, but the 
standard deviation reduces 12--24\% with IMG and IMG-MPS configurations 
(Figure~\ref{fig:endoscopy_line_stddev_x86_all} in Appendix) compared 
to baselines. IMG and IMG-MPS configurations overall perform better across 
all the determinism metrics, as illustrated in Figure~\ref{fig:endoscopy_x86_all}.
Importantly, IMG, IMG-MPS, and IMG-MPS-Pin achieve maximum E2E 
latencies below 50 ms for up to three concurrent endoscopy applications. This 
implies that \textit{a single x86 workstation with two GPUs like A4000 and 
A6000 can run up to three such applications without exceeding 50 ms E2E 
latency}, obviating the need for dedicated workstation per AI application and 
reducing cost, power consumption, and physical footprint in healthcare 
facilities and hospitals without compromising performance predictability and 
safety.

\textit{IMG-MPS improves the maximum E2E latency by 21-30\%, standard 
deviation by 17-24\%, latency distribution tail by 21--47\% and flatness by 
17-25\%}, on average, compared to single GPU setups. 
IMG and IMG-MPS with dual A4000s offer better performance than a single 
A6000, while improving energy usage and reducing cost by over 50\%.
IMG and IMG-MPS yield comparable performance across different metrics, 
underscoring the benefits of isolating compute and graphics tasks on distinct 
GPUs. Nevertheless, IMG-MPS surpasses IMG in reducing the maximum 
latency, latency distribution tail, and flatness by an average of 7\%, 7\%, 
and 8\%, respectively. This performance gain indicates the importance of MPS 
partitioning, even with workload isolation on separate GPUs.

The latency distribution tail in Figure~\ref{fig:endoscopy_tail_x86_all} is below 
75ms for IMG-MPS and IMG-MPS-Pin configurations, demonstrating the 
mitigation of outlier risks. Furthermore, latency 
distribution flatness is constrained to less than 10 ms for IMG-MPS (see 
Figure~\ref{fig:endoscopy_flatness_x86_all}), indicating consistent most-observed 
latencies. These metrics collectively exhibit a holistically greater 
predictable E2E latency for its full distribution of observed values.

More instances raise the maximum latency for IMG-MPS, although at a lower rate, 
due to shared resource contention~\cite{yandrofski2022making}, whose 
mitigation is subject to future work~\cite{ausavarungnirun2018mask}.
Single GPU CUDA MPS partitioning results are not shown here (see 
Section~\ref{sec:cuda_mps_performance}), as it is less 
effective than IMG, IMG-MPS and IMG-MPS-Pin, due to overhead from multiplexing 
graphics context with multiple compute contexts on a GPU outweighing the 
benefits of isolating compute workloads on a single GPU. Pairing MPS 
partitioning with multi-GPU workload isolation results in more performance 
benefits.

\begin{figure*}[b!]
	\centering
	\footnotesize
	\vspace{-14pt}
	\begin{minipage}[t]{0.33\textwidth}
		\centering
		\includegraphics[width=\textwidth]{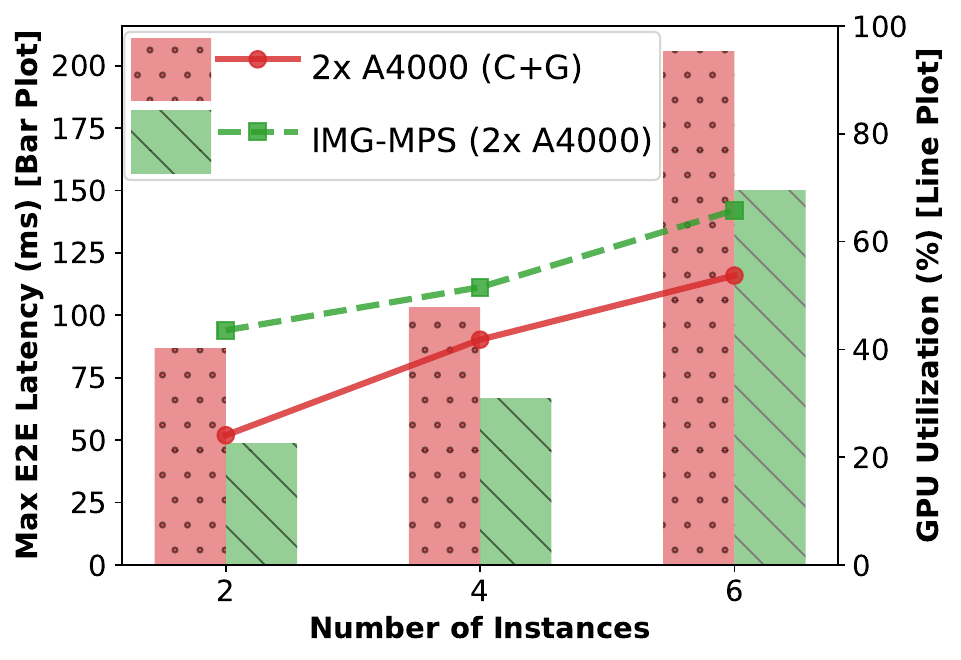}
		\vspace{-10pt}
		\caption{Maximum E2E Latency and GPU Utilization with two A4000s}
		\label{fig:endoscopy_a4000_max_x86_disp}
	\end{minipage}
	\hfill
	\begin{minipage}[t]{0.32\textwidth}
		\includegraphics[width=\textwidth]{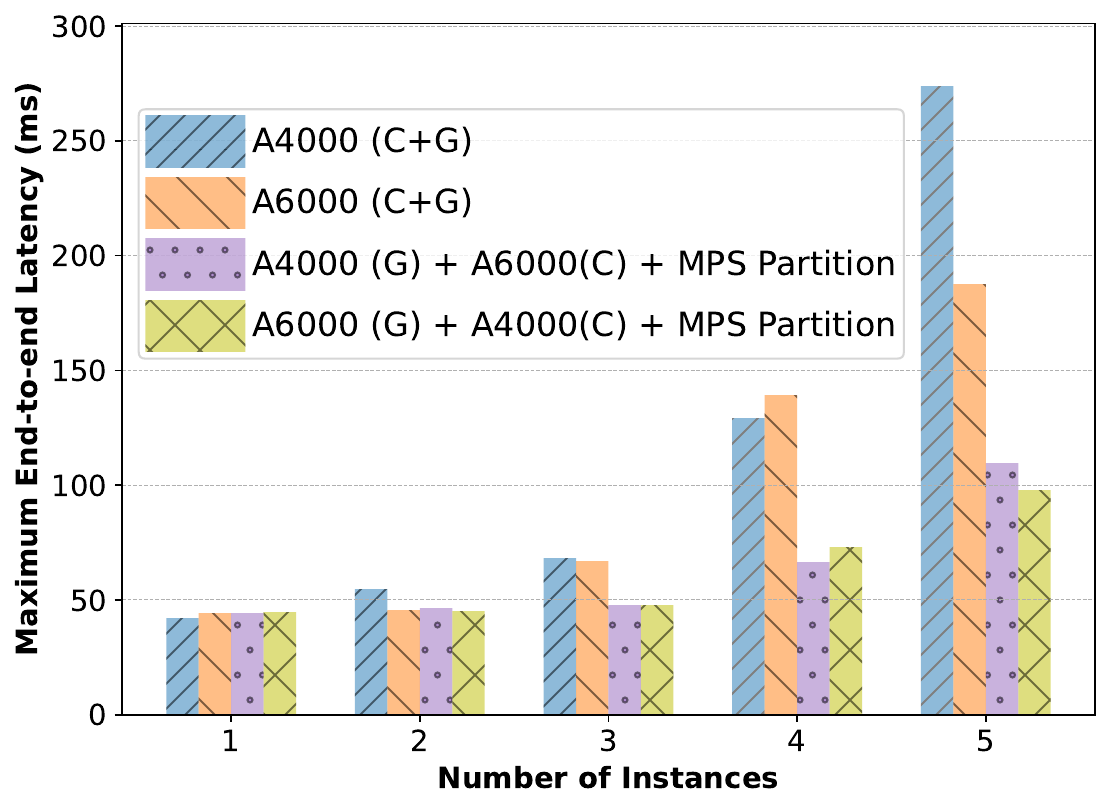}
		\vspace{-10pt}
		\caption{Maximum E2E Latency for different GPU role combinations}
		\label{fig:endoscopy_max_x86_disp}
	\end{minipage}
	\hfill
	\begin{minipage}[t]{0.32\textwidth}
		\includegraphics[width=\columnwidth]{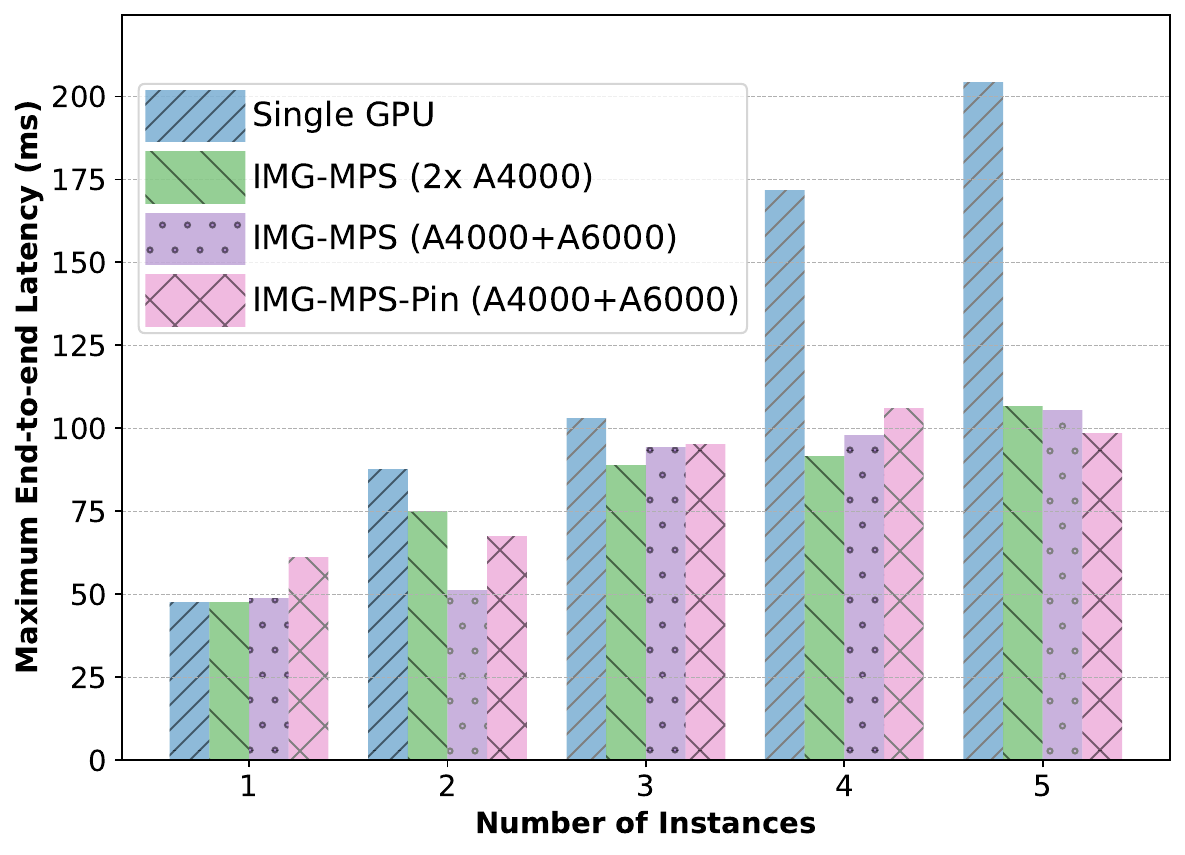}
		\vspace{-10pt}
		\caption{Maximum E2E Latency in Ultrasound Segmentation}
		\label{fig:ultrasound_max_x86_all}
	\end{minipage}
\end{figure*}

\subsubsection*{Impact of CPU Affinity}
Figure~\ref{fig:endoscopy_x86_all} reveals marginal advantages of CPU 
pinning. While some metrics, like maximum E2E latency and latency 
distribution tail, improve in certain cases, others remain static. Given the 
GPU-intensive nature of 
medical AI workloads, CPU affinity yields little
performance gains. Although certification and safety guidelines often 
recommend CPU affinity to mitigate interference, its efficacy in improving 
determinism in medical AI contexts is a bit limited. Nevertheless, CPU 
affinity does not impair performance and is usable for regulatory purposes.

\vspace{-4pt}
\subsection{Do More GPUs Lead to Better Determinism?}
\label{sec:increased_gpu_eval}
Earlier sections showed consistent improvement in performance determinism with 
multi-GPU configurations. In this section, we demonstrate that the computational 
power of multi-GPUs is not the driver behind this performance gain.

We conducted an experiment using two RTX A4000 GPUs in two configurations: 
\textbf{1) Two A4000s (C+G):} Both GPUs are used for compute (C) and graphics (G) 
tasks, and each GPU is connected to a display monitor, \textbf{2) our 
design - A4000 (C) + A4000(G) + MPS Partition:} One A4000 is 
for compute tasks, the other is for graphics, with MPS partitioning enabled at a 
15\% thread percentage, making sure that the sum of maximum active thread 
percentage does not exceed 100\% for up to 6 concurrent applications.
\textit{Figure~\ref{fig:endoscopy_a4000_max_x86_disp} reveals a 35\% average 
reduction in maximum latency and a 42\% increase in GPU utilization with our 
optimized configuration.} These results indicate that our load-balancing design 
is instrumental in achieving predictable latency in a multi-GPU configuration.

Another experiment evaluated the A4000 and A6000 GPUs in four different 
configurations: \textbf{1)} one A4000 for both compute (C) and graphics (G), 
\textbf{2)} one A6000 for both C and G, and \textbf{our design}: \textbf{3)} 
A4000 for G and A6000 for C, 
\textbf{4)} A6000 for G and A4000 for G. 
In Figure~\ref{fig:endoscopy_max_x86_disp}, while our optimized configurations 3 
and 4 do outperform the single-GPU setup like before, the key observation is 
negligible performance difference between the two of our configurations. This 
highlights the effectiveness of our system design, irrespective of the
individual compute capacities of the GPUs.

\vspace{-4pt}
\subsection{Other Applications}
\vspace{-2pt}


\begin{figure*}[t]
	\centering
	\footnotesize
	\begin{subfigure}[t]{0.32\textwidth}
		\includegraphics[width=\linewidth]{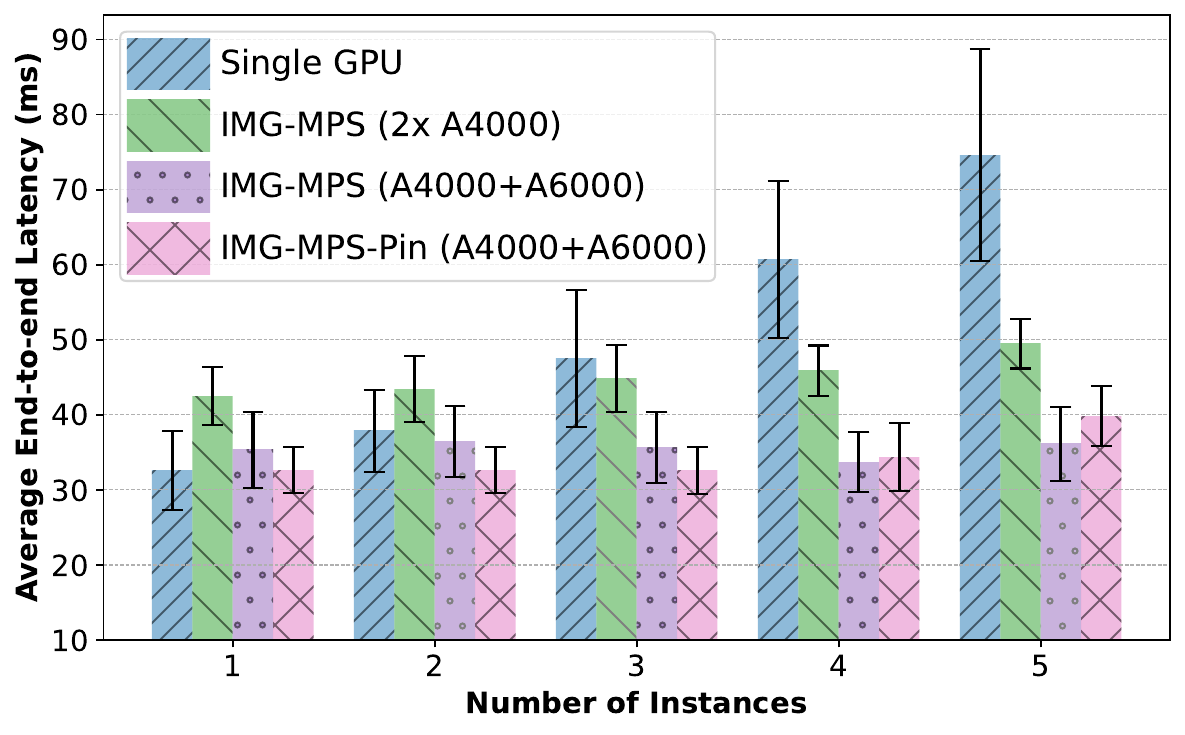}
		\caption{Average End-to-end Latency}
		\label{fig:multiai_ultrasound_avg_x86_all}
	\end{subfigure}
	\hfill
	\begin{subfigure}[t]{0.32\textwidth}
		\includegraphics[width=\linewidth]{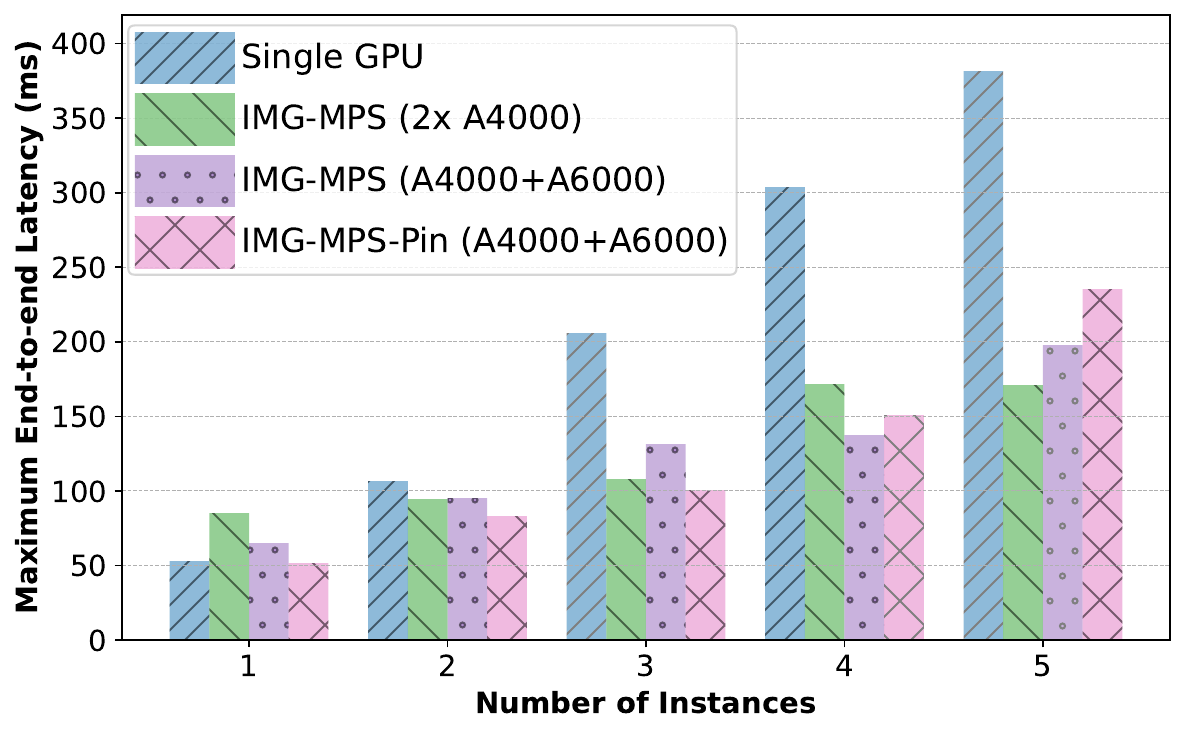}
		\caption{Maximum End-to-end Latency}
		\label{fig:multiai_ultrasound_max_x86_all}
	\end{subfigure}
	\hfill
%
	\begin{subfigure}[t]{0.32\textwidth}
		\includegraphics[width=\linewidth]{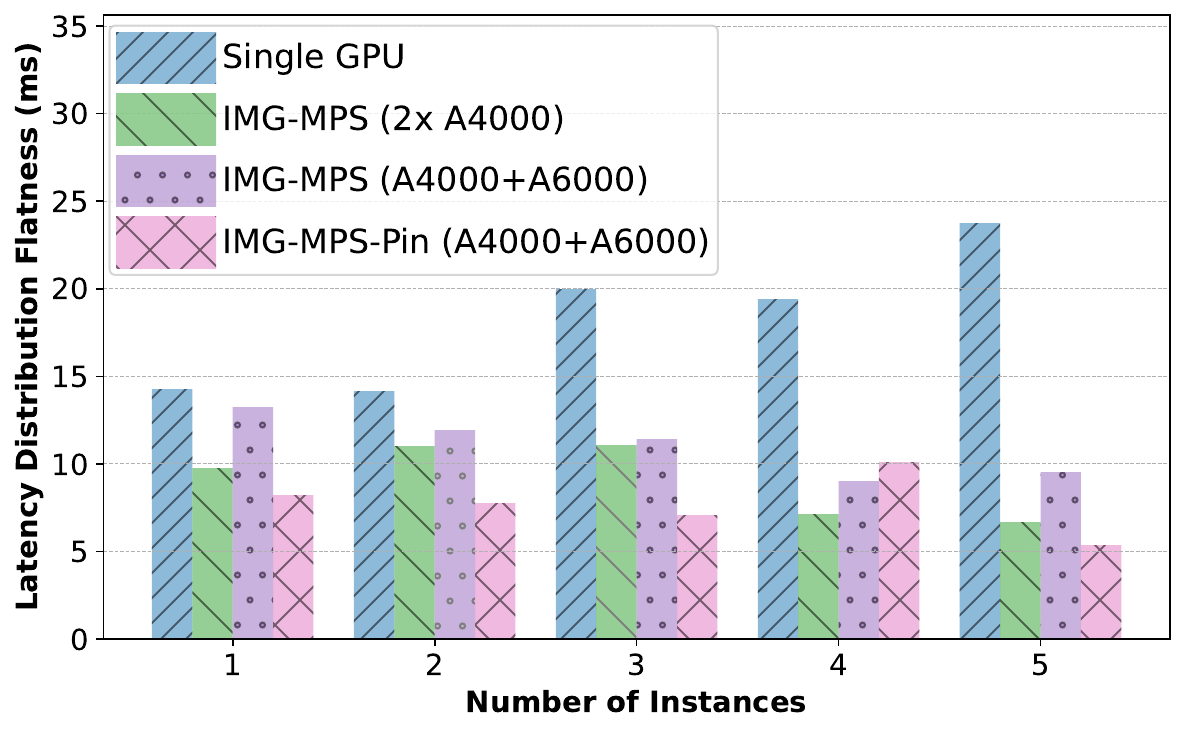}
		\caption{Latency Distribution Flatness}
		\label{fig:multiai_ultrasound_flatness_x86_all}
	\end{subfigure}
	\vspace{-4pt}
	\caption{Multi-AI Ultrasound Application Performance}
	\label{fig:multiai_ultrasound_x86_all}
	\vspace{-18pt}
\end{figure*}

Other \holoscan{} applications, such as ultrasound bone 
scoliosis segmentation and multi-AI cardiac ultrasound~\cite{holohub}, also 
show performance improvements with the IMG-MPS and IMG-MPS-Pin
configurations.
The ultrasound segmentation performs spinal segmentation with a TensorRT
inference model. Its data flow pipeline is similar to the endoscopy 
application, leading to comparable results (see 
Figure~\ref{fig:ultrasound_max_x86_all}).

Figure~\ref{fig:multiai_ultrasound_avg_x86_all} for the multi-AI ultrasound 
shows that the average E2E latency increases with more instances, unlike the 
other applications. As described 
in Section~\ref{sec:multiai_ultrasound}, this application executes three 
parallel inference models, resulting in dramatically elevated GPU workload, 
and consequently, increased GPU resource contention. 
Our design improves both the mean and maximum E2E latency, respectively, by 
4--27\% and 19--33\%. Furthermore, 
Figure~\ref{fig:multiai_ultrasound_flatness_x86_all} 
captures 36--56\% average reduction in flatness using our techniques, peaking at 
56\% improvement through CPU affinity. This optimization minimizes extraneous 
Linux scheduling overhead among the three inference threads in this application, 
as the inferences are time-sliced in parallel CUDA streams on the GPU.

\vspace{-4pt}
\subsection{Heterogeneous Applications}
We revisit the example in Section~\ref{sec:intro} where two different applications 
(endoscopy tool tracking and ultrasound segmentation) are concurrently executed. 
Table~\ref{tab:endo_ultra_together} show that our methods result in a 
maximum E2E latency reduction of 22--32\%. This enables 
manufacturers to consolidate multiple AI applications on a single workstation, 
making devices like endoscopic ultrasound (EUS)~\cite{intuitive_ion,olympus_eus} 
more predictable, safe and affordable. Our techniques also offer a foundational 
step towards multi-function medical device development~\cite{food2021multiple}.
\begin{table}[ht]
	\centering
	\renewcommand{\arraystretch}{1.2} 
	\footnotesize
	\vspace{-4pt}
	\caption{Maximum E2E Latency (ms) for concurrent Endoscopy Tool 
		Tracking and Ultrasound Segmentation}
	\label{tab:endo_ultra_together}
	\vspace{-4pt}
	\begin{tabularx}{\columnwidth}{|p{2.9cm}|p{1cm}|X|X|}\hline
		\multirow{2}{*}{\textbf{Application}} & \multirow{2}{1cm}{\textbf{One 
				A4000}} & \textbf{One A4000 + MPS} & \textbf{Two A4000s + IMG-MPS}
		\\\hline
		Endoscopy Tool Tracking & 133.76  & 
		\multicolumn{1}{c|}{\textbf{\color{white}\cellcolor{nvidiacolor}91.24}}& 
		\multicolumn{1}{c|}{\textbf{\color{white}\cellcolor{nvidiacolor}90.57}}\\\hline
		Ultrasound Segmentation & 80.61 & 
		\multicolumn{1}{c|}{\textbf{\color{white}\cellcolor{nvidiacolor}62.49}} & 
		\multicolumn{1}{c|}{\textbf{\color{white}\cellcolor{nvidiacolor}55.23}} 
		\\\hline
	\end{tabularx}
	\vspace{-8pt}
\end{table}

\vspace{-4pt}
\subsection{Discussion}
Based on our empirical study, we suggest the following design guidelines for 
managing heterogeneous GPU workloads that include equally important compute and 
graphics tasks:
\begin{itemize}[leftmargin=!,labelindent=1pt,align=left]
	\item In scenarios where external constraints prohibit the use of multiple 
	GPUs - due to cost, energy, or other factors - CUDA MPS could be 
	utilized to partition a single GPU, after a thorough resource profiling of the 
	involved applications. However, benefits only with this setup are limited, as 
	shown in Section~\ref{sec:cuda_mps_performance}, due to contention between 
	graphics and multiple compute contexts.
	
	\item Deploying distinct GPUs for compute and graphics tasks greatly enhances 
	performance predictability. For cost considerations, a less powerful GPU may 
	handle graphics if it meets the application's memory needs. For instance, 
	an IGX Orin with future support for 
	inter-operable iGPU and discrete GPU would be a suitable platform.
\end{itemize}

\vspace{-4pt}
\section{Related Work}
\label{sec:related}
Despite the rapid integration of AI technologies into medical devices, ensuring 
deterministic performance in these systems has received little attention.
Existing endoscopy and colonoscopy AI application studies emphasize primarily 
live video processing and accuracy, neglecting end-to-end
performance metrics~\cite{mori2018real,luo2019real}. This work 
addresses the gap by investigating systems techniques to enhance 
time-critical characteristics of medical AI applications.

Prior research in real-time systems has explored priority-aware GPU 
task and interrupt scheduling in time-slicing 
mode~\cite{kato2011timegraph,elliott2012building,kato2012gdev,elliott2013gpusync,tanasic2014enabling,wang2016simultaneous,rossbach2011ptask,capodieci2018deadline,jain2019fractional}.
Our work, however, focuses on the spatial isolation of heterogeneous workloads 
for greater predictability. Inference latency has also been studied and optimized 
in the cloud computing 
context~\cite{gujarati2020serving,romero2021infaas,choi2022serving,ng2023paella,chow2023krisp,dhakal2020gslice},
but these works neglect latency as a safety-critical metric.
Importantly, none of the previous works considered heterogeneous GPU workloads 
involving equally important compute and graphics tasks.

Our approach employs CUDA MPS and heterogeneous workload partitioning across 
multiple GPUs to achieve greater latency predictability. Unlike 
OS-based and device driver 
solutions~\cite{kato2011timegraph,kato2012gdev,rossbach2011ptask,sinha2022towards,sinha2021towards,elliott2013gpusync,jain2019fractional,capodieci2018deadline},
our methodology is applicable to NVIDIA, AMD and other GPUs without requiring 
system software and driver modifications, making it particularly viable for the 
medical industry, which usually mandates long-term (3--10 years) software 
maintenance. Dhakal \textit{et al.} explored CUDA MPS 
for inference performance but did not consider E2E latency 
determinism~\cite{dhakal2020gslice}. 
Zou et al. investigated a theoretical model of 
MPS and simulated task schedulability~\cite{zou2023rtgpu}. 
Our empirical study, in contrast, provides actionable design insights for 
predictable performance. Finally, we use the \holoscan{} 
SDK~\cite{holoscan_sdk}, a novel AI computing platform that resembles the 
programming model of real-time frameworks~\cite{sinha2022solution}
like ROS~\cite{quigley2009ros} and Arduino~\cite{arduino}, 
because it has extensive support for GPUs in the edge-computing domains, 
especially suitable for medical AI applications.


\vspace{-2pt}
\section{Conclusion and Future Work}
\label{sec:conclusion}
This paper presents GPU workload partitioning and load-balancing techniques to 
improve end-to-end latency predictability in medical AI applications using the 
\nvidia{} \holoscan{} platform. By leveraging CUDA MPS partitioning and multi-GPU 
configurations for hardware isolation between compute and graphics workloads, 
we demonstrate significant gains across various latency determinism
metrics against single- and multi-GPU baselines. We expect to motivate 
further 
performance studies involving medical AI workloads. 

Although our study focuses on 
\holoscan{}, it could be used as a design guide for other edge-computing 
systems with concurrent and heterogeneous GPU workloads. Future endeavors 
will investigate solutions for more predictable 
performance on single GPU platforms. Potential benefits of hardware 
virtualization technologies will also be considered in terms of 
predictability, safety, and cost-effectiveness.

\vspace{-4pt}
\bibliographystyle{IEEEtranS}
\bibliography{references}

\clearpage
\section{Appendix}
\label{sec:appendix}

In this section, we include extra results from our experiments for 
clarification.

\begin{figure}[h]
	\centering
	\footnotesize
	\vspace{-8pt}
	\includegraphics[width=1.0\columnwidth]{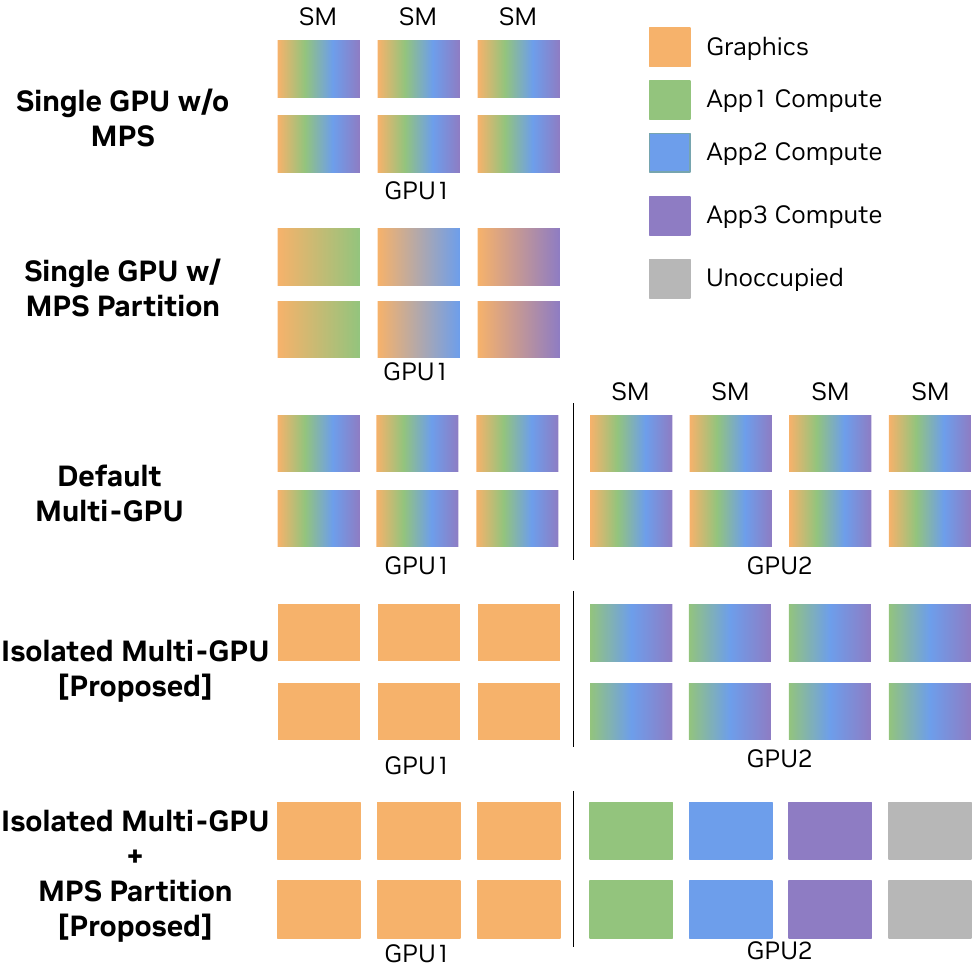}
	\caption{Heterogeneous GPU Workload Partitioning}
	\label{fig:strategy_overview}
	\vspace{-6pt}
\end{figure}

Figure~\ref{fig:strategy_overview} shows the difference of workload 
distribution with different solutions including the techniques introduced in this 
paper. The compute of 3 applications and graphics are color-coded. We imagine 
that every application needs 2 SMs for their compute.
In a single GPU without MPS, graphics and compute of all applications share 
all the SMs. When MPS is enabled on single GPU, compute workload of an 
application is reserved for a 
number of SMs, but graphics workload is shared across all the SMs. Default 
multi-GPU setup is similar to single GPU but with two GPUs. In our proposed 
workload-isolated multi-GPU, graphics is confined to GPU1, but compute of all 
applications still share all the SMs. When MPS is enabled on the workload-isolated 
multi-GPU setup, graphics is isolated to GPU1, and compute workload of every app 
is restricted to a dedicated set of SMs in GPU2, enabling predictable performance 
for AI applications with real-time visualization capabilities. In this 
configuration, the unoccupied SMs could be opportunistically used by the GPU as 
necessary.

\begin{figure}[!ht]
	\centering
	\footnotesize
	\begin{minipage}[b]{0.48\columnwidth}
		\centering
		\includegraphics[width=1.0\textwidth]{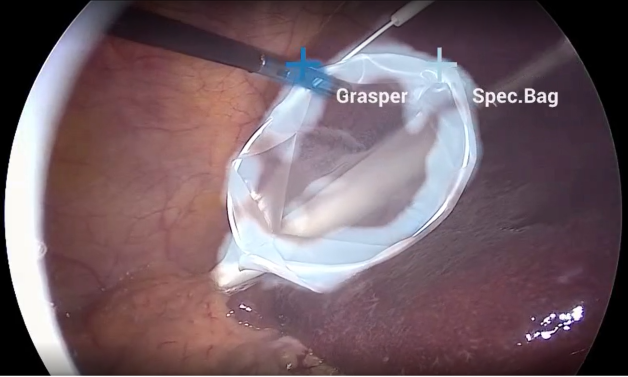}
		\caption{Endoscopy Tool Tracking Sample}
		\label{fig:endoscopy_output}
	\end{minipage}\hfill
	\begin{minipage}[t]{0.48\columnwidth}
		\centering
		\includegraphics[width=1.0\textwidth]{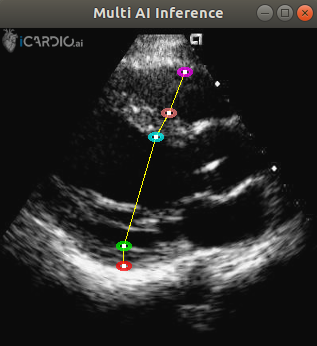}
		\caption{Multi-AI Ultrasound Sample}
		\label{fig:multiai_icardio_output}
	\end{minipage}
	\vspace{-16pt}
\end{figure}

Figure~\ref{fig:endoscopy_output} and Figure~\ref{fig:multiai_icardio_output} show 
sample outputs from the Endoscopy Tool Tracking and Multi-AI ultrasound 
applications described in Section~\ref{sec:medical}.

\subsection{Additional Evaluation}
We report that for most comparisons between different configurations, we have 
observed a Z-score greater than 1.96 with a two-tailed Mann-Whitney U test with 
significance level 0.05, indicating the latency distribution difference. 
However, we do not include this data, as we have analyzed and quantified 
different components of latency distribution for an in-depth view on 
end-to-end latency predictability.

\subsubsection{Impact of Active Thread Percentage in CUDA MPS}
To quantify the effects of active thread percentage value in CUDA MPS 
partitioning, we varied the active thread allocation per MPS client. The 
results, displayed in Table~\ref{tab:endoscopy_mps_variation}, reveal that a 15\% 
allocation counter-intuitively yields better maximum E2E latency and latency 
distribution tail for five instances. This improvement is due to the GPU 
scheduler's dynamic optimization of unallocated threads.
The results underscore the importance of 
application-specific GPU resource profiling. Over-allocation could offer stable 
performance until GPU saturation, at which point it can negatively affect 
co-running processes, as seen with 25\% allocation 
(5x25=125\%). Thus, reserving a few GPU threads for dynamic allocation is 
recommended in CUDA MPS configurations.
\begin{table}[h]
	\centering
	\renewcommand{\arraystretch}{1.2} 
	\footnotesize
	\vspace{-6pt}
	\caption{Impact on E2E Latency for Variation on Active Thread 
		Percentage of CUDA MPS for 5 Instances}
	\label{tab:endoscopy_mps_variation}
	\vspace{-4pt}
	\begin{tabularx}{\columnwidth}{|c|X|X|X|}\hline
		\textbf{Active Thread Percentage} & \textbf{Maximum} & 
		\textbf{Tail} & 
		\textbf{Flatness}\\\hline
		15\% & 94.78 & 56.55 & 8.28 \\\hline
		20\% & 109.58 & 70.82 & 9.28 \\\hline
		25\% & 124.49 & 85.89 & 9.19 \\\hline
	\end{tabularx}
	\vspace{-12pt}
\end{table}
\begin{figure}[b]
	\centering
	\footnotesize
	\vspace{-12pt}
	\includegraphics[width=0.8\columnwidth]{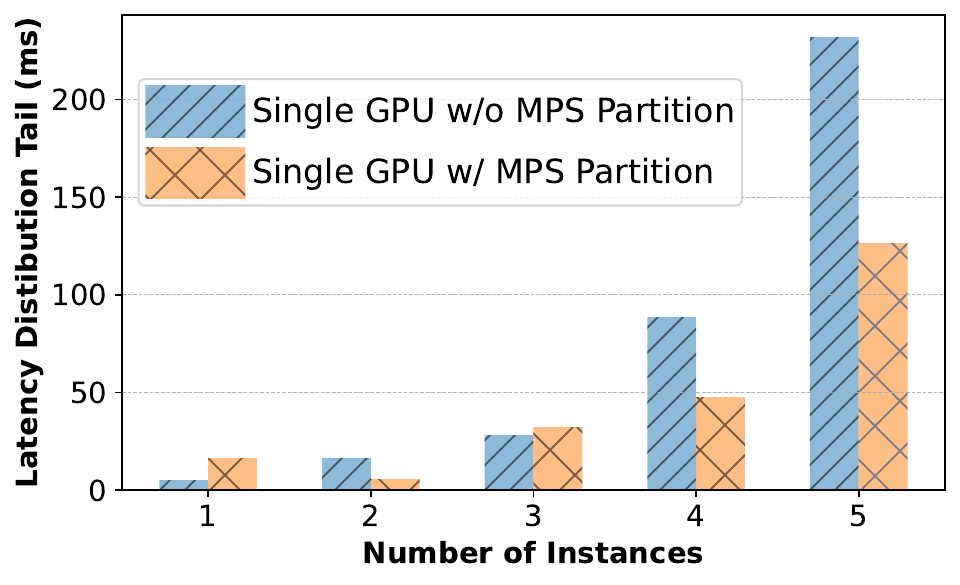}
	\caption{Latency Distribution Tail against Single GPU w/o and w/ CUDA MPS 
		Partition for Endoscopy}
	\label{fig:endoscopy_tail_x86_mps}
\end{figure}
\subsection{Additional Experimental Results}
\begin{figure*}[t]
	\centering
	\begin{minipage}[t]{0.32\textwidth}
		\centering
		\includegraphics[width=\columnwidth]{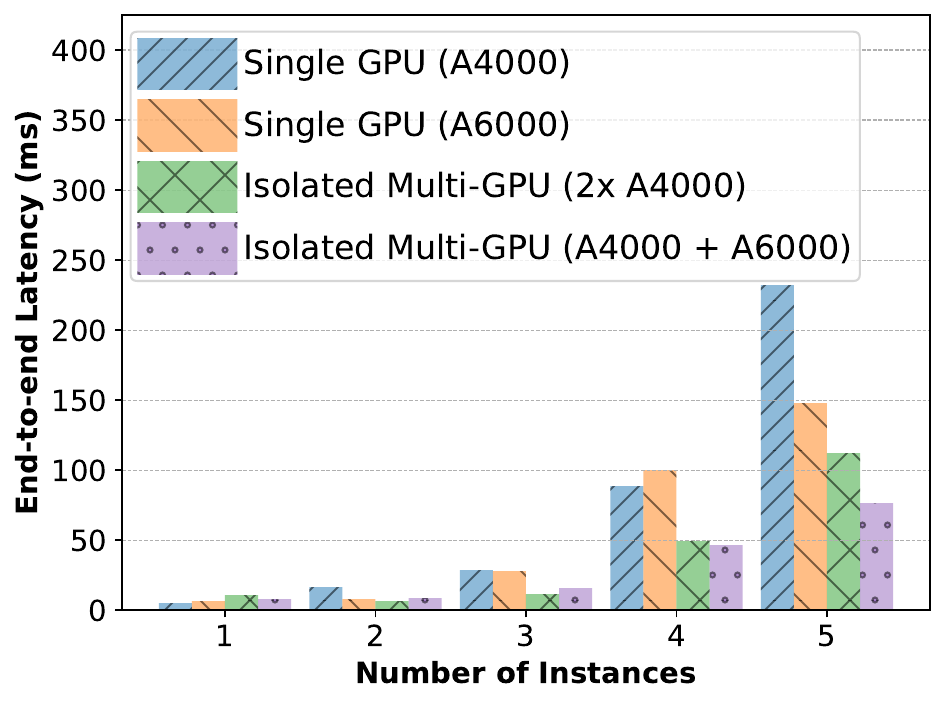}
		\caption{Latency Distribution Tail with Endoscopy Tool Tracking for 
			Single GPU vs. Multi-GPU Configurations}
		\label{fig:isolated_gpu_endoscopy_x86_tail}
	\end{minipage}\hfill
	\begin{minipage}[t]{0.32\textwidth}
		\centering
		\includegraphics[width=\textwidth]{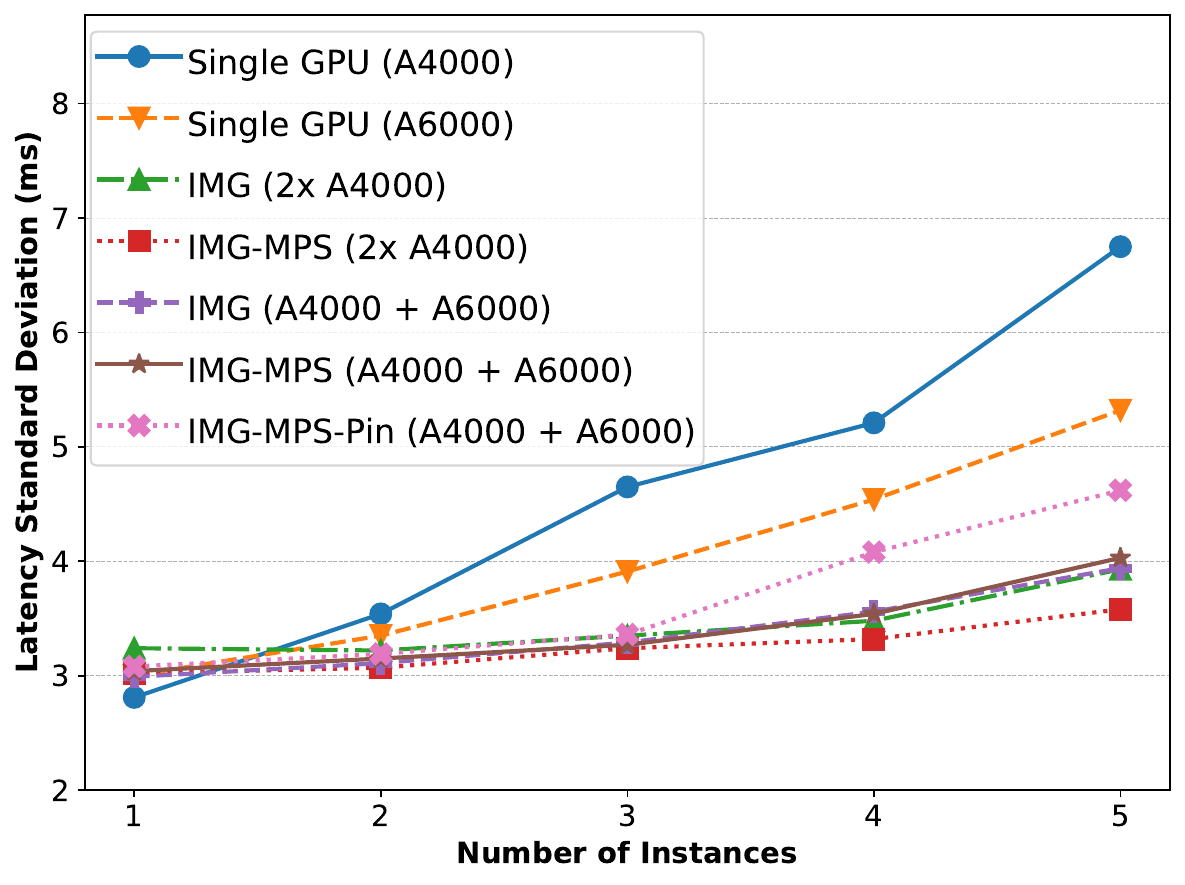}
		\caption{Standard Deviation of Endoscopy Tool Tracking Latency for All 
			Optimizations}
		\label{fig:endoscopy_line_stddev_x86_all}
	\end{minipage}\hfill
	\begin{minipage}[t]{0.32\textwidth}
		\centering
		\includegraphics[width=\columnwidth]{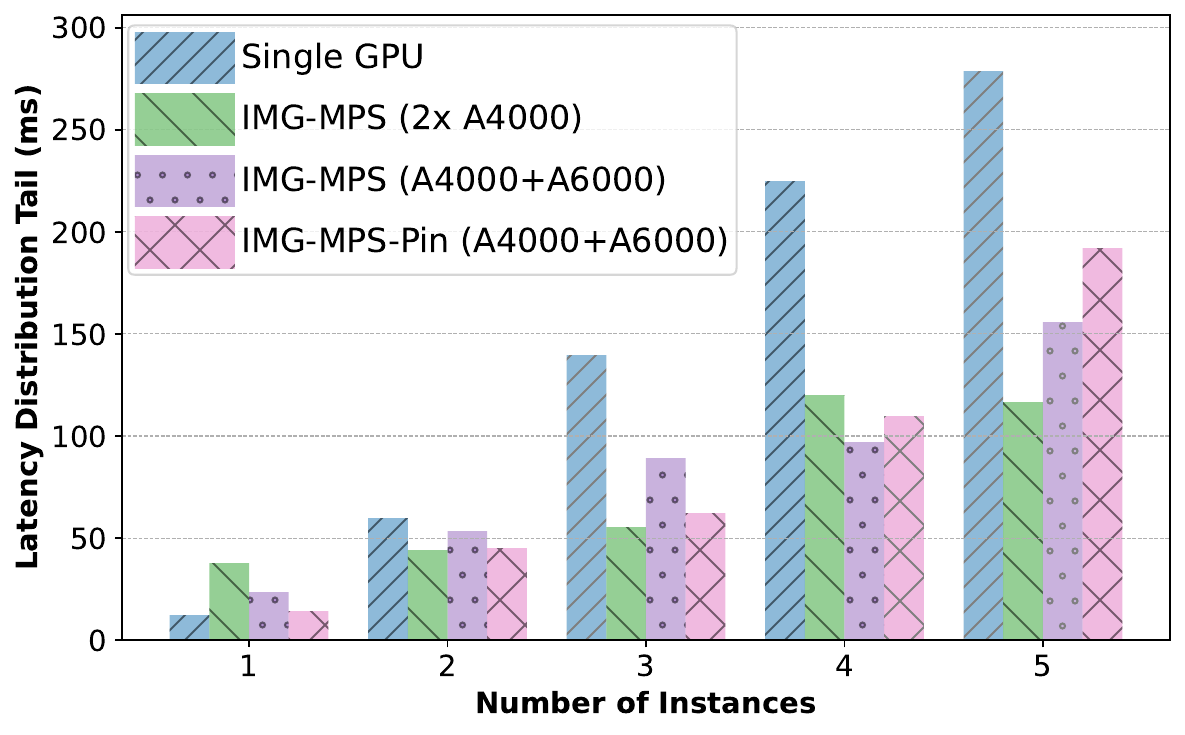}
		\caption{Latency Distribution Tail for Multi-AI Ultrasound}
		\label{fig:multiai_ultrasound_tail_x86_all}
	\end{minipage}
\end{figure*}

Figure~\ref{fig:endoscopy_tail_x86_mps} shows the latency distribution tail for 
the endoscopy tool tracking application against single GPU without and with CUDA 
MPS partitions, whose explanation is already covered in 
Section~\ref{sec:cuda_mps_performance}.
Figure~\ref{fig:isolated_gpu_endoscopy_x86_tail} depicts that workload separation 
in distinct GPUs improves latency distribution tail. 
Figure~\ref{fig:endoscopy_line_stddev_x86_all} shows the standard deviation 
with all optimizations for the endoscopy tool tracking application.

\begin{figure*}[t]
	\centering
	\footnotesize
	\begin{subfigure}[b]{0.32\textwidth}
		\includegraphics[width=\columnwidth]{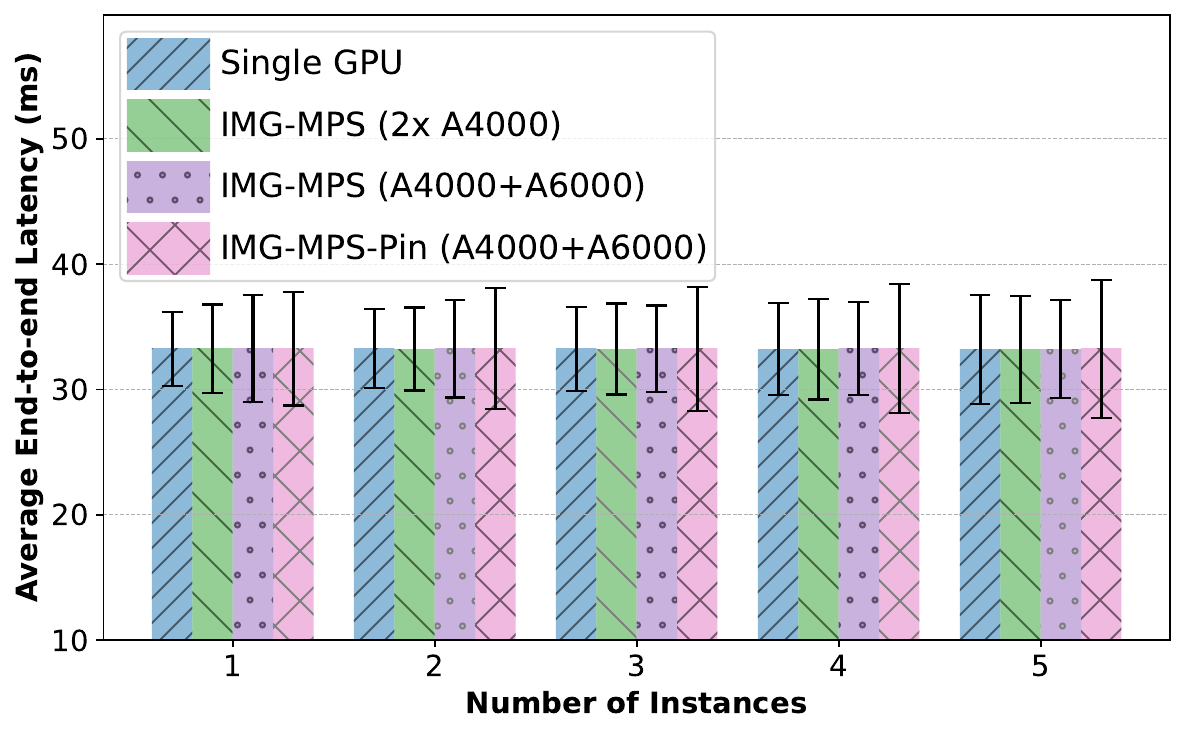}
		\caption{Average E2E Latency}
		\label{fig:ultrasound_avg_x86_all}
	\end{subfigure}\hfill
	\begin{subfigure}[b]{0.32\textwidth}
		\includegraphics[width=\columnwidth]{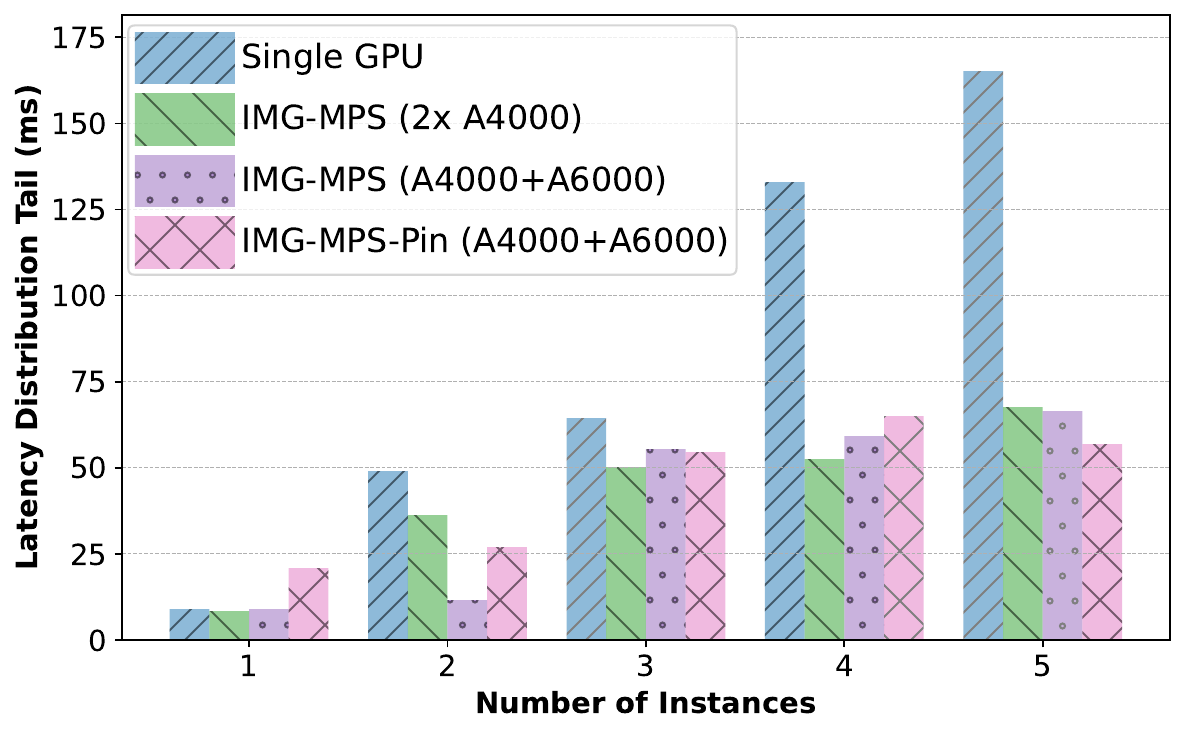}
		\caption{Latency Distribution Tail}
		\label{fig:ultrasound_tail_x86_all}
	\end{subfigure}\hfill
	\begin{subfigure}[b]{0.32\textwidth}
		\includegraphics[width=\columnwidth]{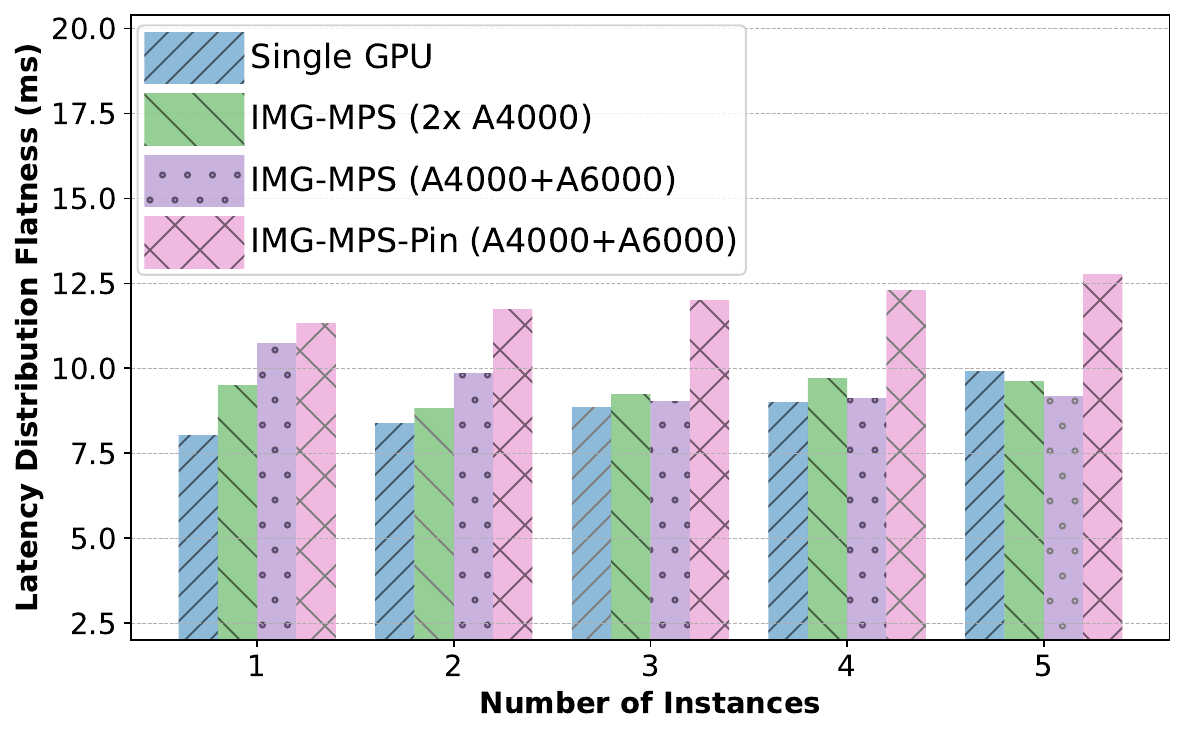}
		\caption{Latency Distribution Flatness}
		\label{fig:ultrasound_flatness_x86_all}
	\end{subfigure}
	\caption{Ultrasound Segmentation Application Performance - rest of the results}
	\label{fig:ultra_rest_results}
\end{figure*}

Our design achieves 10--29\% reduction in latency distribution tail for multi-AI 
ultrasound application, illustrated in 
Figure~\ref{fig:multiai_ultrasound_tail_x86_all}.
Figure~\ref{fig:ultra_rest_results} demonstrates better standard deviation and 
latency distribution tail with our proposed design choices in the 
ultrasound bone scoliosis segmentation application, while the flatness in 
Figure~\ref{fig:ultrasound_flatness_x86_all} is comparable and stable between the 
setups.

\balance

\end{document}